\documentclass[journal]{new-aiaa}
\usepackage{subfigure}
\usepackage{xcolor}
\usepackage{caption}
\usepackage{pstool}
\usepackage{bm}
\usepackage{graphicx}
\usepackage{amsmath}
\usepackage{amsfonts}
\graphicspath{{./figures/}}
\usepackage{siunitx}
\usepackage[version=4]{mhchem}
\usepackage{longtable,tabularx}
\usepackage{comment}

\DeclareMathOperator*{\argmin}{arg\,min}

\title{Random Finite Set Theory and Centralized Control of Large Collaborative Swarms\footnote{this paper is an extension of the paper published in the Proceedings of 2018 American Control Conference, Doerr, Bryce, and Richard Linares. ``Control of Large Swarms via Random Finite Set Theory.'' In 2018 Annual American Control Conference (ACC), pp. 2904-2909. IEEE, 2018.}}

\author{Bryce Doerr\footnote{Postdoctoral Fellow, Department of Aeronautics and Astronautics. Email: {\tt\small bdoerr@mit.edu}, AIAA Member.} and Richard Linares\footnote{Charles Stark Draper Assistant Professor, Department of Aeronautics and Astronautics. Email: {\tt\small linaresr@mit.edu}, Senior AIAA Member.}}
\affil{Massachusetts Institute of Technology, Cambridge, MA, 02139}
\author{Pingping Zhu \footnote{Assistant Professor, Department of Computer Science and Electrical Engineering. Email: {\tt\small zhup@marshall.edu}, AIAA Member.}}
\affil{Marshall University, Huntington, WV, 25755}
\author{Silvia Ferrari \footnote{John Brancaccio Professor, Sibley School of Mechanical and Aerospace Engineering. Email: {\tt\small ferrari@cornell.edu}, AIAA Senior Member.}}
\affil{Cornell University, Ithaca, NY, 14853}

\begin{document}

\maketitle

\begin{abstract}


\textcolor{black}{Controlling large swarms of robotic agents presents many challenges including, but not limited to, computational complexity due to a large number of agents, uncertainty in the functionality of each agent in the swarm, and uncertainty in the swarm's configuration. This work generalizes the swarm state using Random Finite Set (RFS) theory and solves a centralized control problem with a Quasi-Newton optimization through the use of Model Predictive Control (MPC) to overcome the aforementioned challenges. This work uses the RFS formulation to control the distribution of agents assuming an unknown or unspecified number of agents. Computationally efficient solutions are also obtained via the MPC version of the Iterative Linear Quadratic Regulator (ILQR), a variant of Differential Dynamic Programming (DDP). Information divergence is used to define the distance between the swarm RFS and the desired swarm configuration }
\textcolor{black}{through the use of the modified $L_2^2$ distance. Simulation results using MPC and ILQR show that the swarm intensity converges to the desired intensity. Additionally, the RFS control formulation is shown to be very flexible in terms of the number of agents in the swarm and configuration of the desired Gaussian mixtures. Lastly, the ILQR and the Gaussian Mixture Probability Hypothesis Density filter are used in conjunction to solve a spacecraft relative motion problem with imperfect information to show the viability of centralized RFS control for this real-world scenario.}


\end{abstract}

\section{INTRODUCTION}
Control of large collaborative networks or swarms is currently an emerging area for controls development. Typically, a swarm network is comprised of tiny robots with limited actuators that perform specific tasks in some collective configuration. For example, the swarm can use its collective effort to grasp or move in a changing environment which can offer more flexibility and redundancy to meet a goal compared to the abilities of a single agent \cite{1}. Specifically in space applications, swarm control of satellites and rovers can be used for the exploration of asteroids and other celestial bodies of interest \cite{2} or areas of assembly and construction on-orbit, including constructing space observatories and space habitats \cite{izzo2005mission}. Swarms involving UAVs have proven to be widely useful in military applications such as search and rescue missions, communication relaying, border patrol, surveillance, and mapping of hostile territory \cite{3}. From these engineering applications, the use of collaborative swarms is an attractive option to meet objectives that require flexibility and redundancy in a changing environment.
For collaborative swarms, several control techniques have been implemented to date. With centralized control, one agent in the swarm computes the overall swarm control and manages the control execution for individual agents allowing it to oversee the other agents' system processes \cite{5}. Unfortunately, centralized control suffers from two main problems. As the number of agents in the swarm increases, the computational workload becomes more expensive \cite{bakule2008decentralized}. This is especially true when the swarm agents are low-cost and are located in an unknown environment \cite{4}. For example, formation control was applied experimentally to 1024 low-cost Kilobot robots which took 12 hours to converge to a specific formation \cite{4}. Additionally, centralized control is not robust against individual agent failures \cite{6}. With a thousand low-cost agents present in a swarm, communication, actuation, and sensing are performed with less reliability. Thus, control formulation must be found which considers the computational performance for control of low-cost agents as well as flexibility during uncertainty.  

Random Finite Set (RFS) formalism provides a generalization of the state space for multi-agent systems which can be used for control \cite{19,28,44}. It is used to solve a stochastic trans-dimensional problem, where the dimension of the state-space is an unknown a priori (unknown number of agents). RFS allows for the probability density function over a collection of state-spaces to be defined, which provides a potential hypothesis for the true number of agents. Then a Bayesian estimation problem is formulated and approximate solutions are used through the Gaussian Mixture Probability Hypothesis Density (GM-PHD) filter \cite{27}. Other than the GM-PHD filter, many extensions around RFS theory have been made using estimation and simultaneous localization and mapping (SLAM) techniques including the Cardinalized Probability Hypothesis Density (CPHD) filter and the Generalized Labeled Multi-Bernoulli (GLMB) filter \cite{19,27,Vo_2006,Vo_2013}.

By using RFS theory to model multi-agent systems, the time-varying number of agents and their states can be jointly estimated from measurement sets including data association uncertainty, clutter, and noise \cite{19,27}. The agents and measurements are modeled as RFSs, and the Probability Hypothesis Density (PHD) filter is used to propagate the estimate forward in time. The RFS model has been used previously with a potential model to describe the temporal evolution of the probabilistic description of a robotic swarm to promote coordination \cite{28}. Other work has developed control for individual agents using the estimated RFS state in a centralized fashion \cite{44}. As an introduction to RFSs, the GM-PHD filter is explored for application to RFS control.



Other models were developed to represent the behavior for swarm agents in space and time including probabilistic swarm guidance and distributed optimal control which has developed efficient decision making for swarm control \cite{9,37,38,39,41}. Probabilistic swarm guidance has been used to enable swarms to converge to target distributions through distributed control \cite{9}. 
Distributed control is defined as the reformulation of the control problem as a set of interdependent subproblems and solving these subproblems \cite{8}. Probabilistic swarm guidance solves issues that involve a large number of agents, also identified as ``computationally complex'', by controlling the swarm density distribution of the agents \cite{9}.
The distributed optimal control method is a method that controls multi-agent systems by modeling the agents as Gaussian mixtures and using an integral cost function that is optimized to the advection equation \cite{37,38,39}. The control laws themselves are determined using potential functions that attract the agent distributions to the desired state and repel the distribution from obstacles \cite{40}. By minimizing the objective function based on distributions using the necessary conditions of optimality, the optimal control law is found using the potential function. The distributed optimal control method was also expanded to use the Kullback-Leibler divergence metric using distributions in the objective function for the use of path planning \cite{41}. This provides a discovery to a whole class of divergence measures of distributions that can provide converging optimal control solutions to multi-agent systems.

Decentralized control has also been implemented in regards to swarm control. Decentralized Model Predictive Control (MPC) was applied to swarms of low-cost spacecraft with limited capabilities for swarm reconfiguration \cite{10}.
The benefit of this solution is that it decentralizes the computation and communication required for the swarm system. 
In \cite{18}, they used decentralized planning and sequential convex programming to control swarms. 
Using sequential convex programming in combination with MPC in real time provided robustness as the agents converged to designated targets. The same authors also used sequential convex programming to do target assignment (mapping of agents to targets) and trajectory generation for varying swarm sizes through time \cite{15}.

\begin{figure}[h]
\begin{centering}
      \includegraphics[width=1\textwidth]{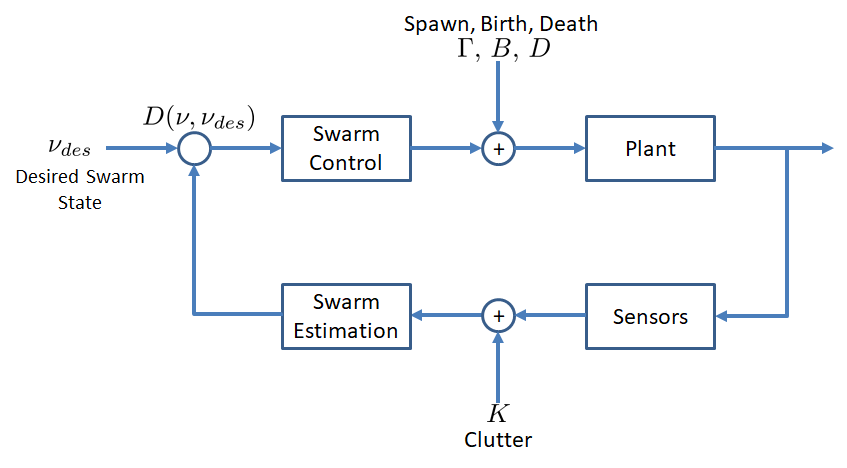}
        \caption{A block diagram of the RFS control and estimation architecture in a closed-loop. }\label{closedloop}
\end{centering}
\end{figure}

\textcolor{black}{The objective of this paper is the formulation of the swarm estimation and control problems using RFS theory. The main contributions of this paper are:}
\begin{enumerate}
	\item The generalization of the state representation using RFS theory for the control of large collaborative swarms under unknown number of agents.
	\item The proposal of new distributional-based distances for the control cost function.  
	\item The ``closing-the-loop'' between RFS control and the PHD filter.
	\item The application of multi-agent estimation and control using RFSs for formation flying of varying number of large collaborative swarms with the inclusion of process and measurement noise.
\end{enumerate}
\textcolor{black}{The first contribution is accomplished by representing the swarm state with a RFS, where RFSs are a collection of agent states, with no ordering between individual agents, that can randomly change through time \cite{19}. For contribution two, several key divergence metrics are considered as control cost functions to drive the overall swarm behavior to a desired configuration. The third contribution is shown in Figure \ref{closedloop}. In Figure \ref{closedloop}, the first moment of the RFS models the current RFS swarm configuration, $\nu$, and the desired RFS swarm configuration is defined by its first moment, $\nu_{des}$. The first moment (or intensity discussed later on) contains information on the number of agents and their states. The PHD filter is used to process measurements from an unknown number of agents with defined spawn ($\Gamma$), birth ($B$), and death ($D$) rates, and the distributional distance-based cost ``closes-the-loop'' for RFS swarm control. Note that the number of agents is not a controlled quantity and in fact this work assumes that the number of agents is unknown and estimated under the RFS formulation. For the last contribution, convergent control solutions through MPC and Differential Dynamic Programming (DDP) are found from GM-PHD filter estimates to advance control methods for swarm applications (e.g. Clohessy-Wiltshire relative motion) which offers improvements in computational efficiency and flexibility to uncertainty. Although the topology underlying the work is centralized, the formulation allows for a statistical model of the swarm and provides improvements in computational efficiency and flexibility (similar to the distributed optimal control methods \cite{9,37,38,39,41}) when compared to traditional centralized methods \cite{4}. Although not presented in this paper, the formulation in general can naturally be extended to decentralized control which is expected to provide decentralized communication and computational efficiency.}

The paper is organized as follows. Section II introduces preliminaries of the RFS theory relevant to this work. Section III presents the RFS-based control problem formulation which is a central contribution of this work. Section IV proposes new distributional distance-based cost functions to form the RFS control problem. Section V discusses the dynamical models that will demonstrate the methods and solutions discussed in the previous section. Section VI discusses relevant simulation results involving the RFS control framework. Section VII provides limitations to the RFS control work presented. Lastly in Section VIII, concluding remarks are provided.

\section{Preliminaries}

The swarm control approach developed in this work makes use of RFS theory. The motivation for the use of RFS theory stems from its application to multi-agent tracking \cite{19,27}. We begin by presenting the single-agent and multi-agent tracking problems. Additionally, the multi-agent RFS-based tracking problem is used in this work to ``close-the-loop” for control of large collaborative swarms.

The discrete dynamics and measurement model for a time-varying, single-agent system between discrete time-steps $k$ and $k+1$ is given by
\begin{subequations}
\begin{equation}
    \mathbf{x}_{k+1}=A_{k}\mathbf{x}_{k}+B_k\mathbf{u}_k+\boldsymbol{\epsilon}_k,
\end{equation}
\begin{equation}
    \mathbf{z}_k=H_k\mathbf{x}_{k}+D_k\mathbf{u}_k+\boldsymbol{\sigma}_k,
\end{equation}
\end{subequations}
where $\mathbf{x}_k$ is the agent state, $A_k$ is the system matrix, $B_k$ is the control input matrix, $\mathbf{u}_k$ is the control input, $\mathbf{z}_k$ is the measurement vector, $H_k$ is the observation matrix, and $D_k$ is the feed-forward matrix. The process noise, $\boldsymbol{\epsilon}_k$, and measurement noise, $\boldsymbol{\sigma}_k$, are zero mean Gaussian noise with variances $\Sigma_{\boldsymbol{\epsilon}}$ and $\Sigma_{\boldsymbol{\sigma}}$, respectively. Note that a single agent, ${\mathbf{x}}_{k}$, is produced from a space $\mathcal{X}\subseteq\mathbb{R}^{d_x}$ where $d_x$ is the agent state vector size. Similarly, the single agent's control input, $\mathbf{u}_k$, is produced from a space $\mathcal{U}\subseteq\mathbb{R}^{d_u}$ where $d_u$ is the agent's control vector size, and ${\bf z}_{k}$ is produced from a space $\mathcal{Z}\subseteq\mathbb{R}^{d_z}$ where $d_z$ is the measurement vector size.
\subsection{Single-Agent Filtering}
To estimate the dynamics for a single-agent system, it is usually assumed that the state space follows a Markov process with a transition density, 
\begin{equation}\label{transmatrix}
f_{k|k-1}\left( {\mathbf{x}}_{k}|{\mathbf{x}}_{k-1}\right),
\end{equation}
which is the probability density for the single-agent system to move through its dynamics from $k-1$ to $k$. For generality, the dynamical system is partially observed as a likelihood function given by
\begin{equation}\label{likelihoodmatrix}
g_k\left( {\bf z}_{k}|{\mathbf{x}}_{k}\right),
\end{equation}
where the likelihood function is a probability density of observing the system by obtaining measurements, ${\bf z}_{k}$. By using the observation information from ${\bf z}_{1:{k}}=\left( {\bf z}_1,\cdots,{\bf z}_{k}\right)$, the posterior density estimate at a time $k$ is determined using the Bayesian recursion given by
\begin{subequations}
\begin{equation}
p_{k|k-1}\left( {\mathbf{x}}_{k}|{\bf z}_{1:k-1}\right)=\int f_{k|k-1}\left( {\mathbf{x}}_{k}|{\mathbf{x}}_{k-1}\right)p_{k-1}\left( {\mathbf{x}}_{k-1}|{\bf z}_{1:k-1}\right)d{\mathbf{x}}_{k-1},
\end{equation}
\begin{equation}
p_{k}\left( {\mathbf{x}}_{k}|{\bf z}_{1:k}\right)=\frac{g_k\left( {\bf z}_{k}|{\bf x}_{k}\right)p_{k|k-1}\left( {\mathbf{x}}_{k}|{\bf z}_{1:k-1}\right)}{\int g_k\left( {\bf z}_{k}|{\mathbf{x}}_{k}\right)p_{k|k-1}\left( {\mathbf{x}}_{k}|{\bf z}_{1:k-1}\right)d\mathbf{x}_k}.
\end{equation}
\end{subequations}
The posterior density contains the measurement update, and the estimate for this single-agent system can be found using a minimum mean squared error method.
\subsection{RFS Formulation}
For the multi-agent tracking problem, a Bayesian recursion through a RFS formulation with discrete-time dynamics is considered \cite{27}. 
This theory addresses the decentralized estimation problem for each agent in the formation.
An $i$th agent in the swarm at time-step $k$ has the challenge of estimating its state configuration ($\mathbf{x}^i_{k}\in \mathcal{X}\subseteq\mathbb{R}^{d_x}$) and designing a control policy to achieve that state. In this work, it is assumed that each agent within the swarm is identical, and using unique identifiers on each agent is unnecessary. Using this theory, the RFS models the uncertainty (i.e. the number of agents and their spatial states) by a random finite set  \cite{27}. The agents in the field may die, survive and move into the next state through dynamics, or appear by spawning or birthing.  The unknown number of agents in the field is denoted by $N_{\text{total}}(k)$ and may be randomly varying at each time-step by the union of the birth $\left(\Gamma_k:\emptyset\rightarrow \left\lbrace {\bf x}^i_{k},{\bf x}^{i+1}_{k},\cdots,{\bf x}^{i+N_{birth(k)}}_{k}\right\rbrace\right)$, spawn $\left(B_{k|k-1}\left( \mathbf{x}^{i}_{k-1}\right):{\bf x}^{i}_{k-1}\rightarrow \left\lbrace {\bf x}^{i}_{k},{\bf x}^{i+1}_{k},\cdots,{\bf x}^{i+N_{spawn(k)}}_{k}\right\rbrace\right)$, and surviving $\left(S_{k|k-1}\left( \mathbf{x}^i_{k-1}\right):{\bf x}^i_{k-1}\rightarrow{\bf x}^i_{k}\right)$ agents. Death is denoted by $D_k\left( \mathbf{x}^i_{k-1}\right):{\bf x}^i_{k-1}\rightarrow\emptyset$. The number of births, $N_{birth(k)}$, and the number of spawns, $N_{spawn(k)}$, are unknown quantities that vary at each time-step. The RFS, $X_k$, that describes the births, spawns, deaths, and surviving agents is given by
\begin{equation}\label{unionstate}
X_k=\left[ \bigcup_{\mathbf{x}^i_{k-1}\in X_{k-1}} S_{k|k-1}\left( \mathbf{x}^i_{k-1}\right)\right]\cup\left[ \bigcup_{\mathbf{x}^i_{k-1}\in X_{k-1}}B_{k|k-1}\left( \mathbf{x}^i_{k-1}\right)\right]\cup \Gamma_k.
\end{equation}
$X_k= \left\{{\bf x}^1_{k},{\bf x}^2_{k},\cdots,{\bf x}^{N_{total(k)}}_{k} \right\}$ denotes a realization of the RFS distribution for agents. The individual RFSs in Eq. \eqref{unionstate} are assumed to be independent from each other. For example, any births that occur at any time-step are independent from any surviving agents.
At any time, $k$, the RFS probability density function can be written as
\begin{equation}\label{rfsstate}
p(X_k=\left\{{\bf x}^1_{k},{\bf x}^2_{k},\cdots,{\bf x}^n_{k}\right\})=p(|X_k|=n)p(\left\{{\bf x}^1_{k},{\bf x}^2_{k},\cdots,{\bf x}^n_{k}\right\}{\mid \lvert X_k\rvert =n)}.
\end{equation}
For a generalized observation process, the agents are either detected $\left(\Theta_k\left( {\bf x}^i_{k}\right):{\bf x}^i_{k}\rightarrow{\bf z}^i_{k}\right)$, or they are not detected $\left(F_k\left( {\bf x}^i_{k}\right):{\bf x}^i_{k}\rightarrow\emptyset\right)$. Clutter or false alarms $\left(K_k: \emptyset\rightarrow \left\lbrace {\bf z}^1_{k},{\bf z}^2_{k},\cdots,{\bf z}^{N_{clutter}}_{k}\right\rbrace\right)$, defined as measurements that do not belong to any agents, are also present in the set of observations. For a time-step $k$, note that ${\bf z}^i_{k}$ is the $i$th measurement obtained from a space $\mathcal{Z}\subseteq\mathbb{R}^{d_z}$. Therefore, RFS of measurements is described by
\begin{equation}\label{unionmeasurement}
Z_k=K_k\cup\left[ \bigcup_{ {\bf x}^i_{k}\in X_k }\Theta_k\left(  {\bf x}^i_{k}\right)\right],
\end{equation}
where the origins of each measurement are not known and unique identifiers are not necessary. Again, the individual RFSs in Eq. \eqref{unionmeasurement} are independent of each other, so measurements and clutter are obtained independently from each other. With $X_k$ and $Z_k$ defined over sets of agents' states and measurements, a Bayesian recursion for multi-state estimation can be applied.

On a similar note, the control sequence is also defined by a RFS in the form $U_k= \left\{{\bf u}^1_{k},{\bf u}^2_{k},\cdots,{\bf u}^{N_{total(k)}}_{k} \right\}$ and a RFS probability density given by
\begin{equation}\label{rfsstate2}
p(U_k=\left\{{\bf u}^1_{k},{\bf u}^2_{k},\cdots,{\bf u}^n_{k}\right\})=p(|U_k|=n)p(\left\{{\bf u}^1_{k},{\bf u}^2_{k},\cdots,{\bf u}^n_{k}\right\}{\mid \lvert U_k\rvert =n)},
\end{equation}
\textcolor{black}{since the realization of agents on the field to be controlled are varying with time, $k$}. Similarly, note that ${\bf u}^i_{k}$ is the $i$th agent's control input obtained from a space $\mathcal{U}\subseteq\mathbb{R}^{d_u}$.

The random finite set formulation of describing multi-agent states and observations can be described very similarly to Eq. \eqref{transmatrix} and \eqref{likelihoodmatrix} for single agent estimation, but the RFS states ($X_k$) and observations ($Z_k$) are used instead. To determine the multi-agent posterior density, a multi-agent Bayes recursion is used given by
\begin{subequations}
\begin{equation}
p_{k|k-1}\left( X_{k}|Z_{1:k-1}\right)=\int f_{k|k-1}\left( X_{k}|X_{k-1}\right)p_{k-1}\left( X_{k-1}|Z_{1:k-1}\right)\mu_s(dX_{k-1}),
\end{equation}
\begin{equation}\label{RFSgenmeasure}
p_{k}\left( X_{k}|Z_{1:k}\right)=\frac{g_k\left( Z_{k}|X_{k}\right)p_{k|k-1}\left( X_{k}|Z_{1:k-1}\right)}{\int g_k\left( Z_{k}|X_{k}\right)p_{k|k-1}\left( X_{k}|Z_{1:k-1}\right)\mu_s(dX_{k})},
\end{equation}
\end{subequations}
where $\mu_s$ is a reference measure on a collection of all finite subsets of state space \cite{27}. From Eq. \eqref{RFSgenmeasure}, the integration occurs over all possible locations of agents residing in the state space as well as their number which becomes a set integral. The recursion thus contains uncertainty in the agent number and location brought by detection uncertainty and measurement noise, respectively. Measurements are not a direct function of the individual agents due to explicitly incorporating clutter into the formulation. Therefore, measurement to agent assignment is not explicitly required in the formulation.
By computing the set integral about all possible number of agents and their states, the recursion can become intractable,  but solutions have been found for a small number of agents using sequential Monte Carlo \cite{35}. Fortunately, a PHD filter approximation provides computational tractability for larger numbers of agents.

\subsection{Probability Hypothesis Density (PHD) Filter}
Instead of propagating the multi-agent posterior density through a multi-agent Bayes recursion, the Probability Hypothesis Density (PHD) filter propagates the posterior intensity function. The nonnegative intensity function, $v({\boldsymbol{\xi}})$, is a first-order statistical moment of the RFS state that represents the probability of finding an agent, represented by a generalized state variable $\boldsymbol{\xi}\in\mathcal{X}$, in a region of state space $\mathcal{S}\subseteq\mathcal{X}$. The estimated number of agents in the region $\mathcal{S}$ is the integral of the intensity function given by
\begin{equation}\label{generalintensityintegral}
\mathbb{E}(|X\cap \mathcal{S}|)=\int_{\mathcal{S}}v(\boldsymbol{\xi})d\boldsymbol{\xi},
\end{equation}
where the expectation represents a RFS $X$ intersecting a region $\mathcal{S}$. This gives the total mass or the number of estimated agents for RFS $X$ in a region $\mathcal{S}$. The local maximum in intensity $v(\boldsymbol{\xi})$ shows the highest concentration of expected number of agents which can be used to determine an estimate for the agents in $X$ at a time-step.

\begin{figure}[!htb]
\begin{centering}
\includegraphics[keepaspectratio, width=.64\textwidth]{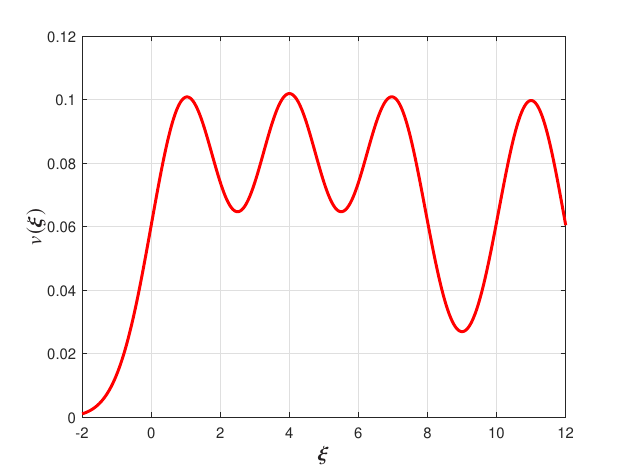}
        \caption{The intensity for a 1D, 4-agent problem using Eq. \eqref{gmexample}.}\label{1Dprob}
\end{centering}
\end{figure}

To further interpret the intensity function, consider a one-dimensional example with four agents located with a mean and covariance of $\mathbf{m}=\{1, 4, 7, 11 \}$ and $P^i=1:i = 1,\dots,4$, respectively. This is a realization of $X$. Assuming the intensity function corresponds to a Gaussian mixture representation given by
\begin{equation}\label{gmexample}
    \nu(\boldsymbol{\xi})=\sum_{i=1}^{N_{total}}w^{(i)}\mathcal{N}\left(\boldsymbol{\xi};{\bf m}^i,P^i\right),
\end{equation}
where each weight $w^{(i)}=1$, $\nu(\boldsymbol{\xi})$ can be plotted against the generalized state $\boldsymbol{\xi}$ given by Figure \ref{1Dprob}.
\textcolor{black}{Note that this example is a specific case in which the number of agents are equal to the number of Gaussian mixtures by assuming $w^{(i)}=1$, while in general, this is not the case. This is further discussed in the Random Finite Set Control Problem Formulation section.} The agent locations (X=\{1, 4, 7, 11 \}) are located at individual maxima of $\nu(\boldsymbol{\xi})$. The integral of $\nu(\boldsymbol{\xi})$ in Eq. \eqref{generalintensityintegral} is
\begin{equation}
\begin{split}
 \mathbb{E}(|X\cap S|)=\int_Sv(\boldsymbol{\xi})d\boldsymbol{\xi}&=\int\mathcal{N}\left(\boldsymbol{\xi};1,1\right)d\boldsymbol{\xi}+\int\mathcal{N}\left(\boldsymbol{\xi};4,1\right)d\boldsymbol{\xi} +\int\mathcal{N}\left(\boldsymbol{\xi};7,1\right)d\boldsymbol{\xi}+\int\mathcal{N}\left(\boldsymbol{\xi};11,1\right)d\boldsymbol{\xi} \\
 &=1\hspace{12pt}+\hspace{12pt}1\hspace{12pt}+\hspace{12pt}1\hspace{12pt}+\hspace{12pt}1\hspace{12pt}=4,
 \end{split}
\end{equation}
which is the total mass or the total number of estimated agents of RFS $X$ in this state space $\mathcal{S}$. It is also noted that the intensity function is not a probability density since the integral over $\boldsymbol{\xi}$ does not generally sum up into one. By estimating a potentially large, single intensity function, $\nu(\boldsymbol{\xi})$, an estimate of the number of agents and their states can be obtained.

A RFS that is fully characterized by their intensity is the Poisson RFS. By assuming the RFS $X$ is Poisson of the form $p(|X|=n)$ and $p(\left\{{\bf x}^1,{\bf x}^2,...,{\bf x}^n\right\}{\mid \lvert X\rvert =n)}$, approximate solutions can be determined by the PHD filter \cite{19,27}. Propagation of the PHD can be determined if the agents are assumed to be independent and identically (i.i.d.) with the cardinality of the agent set that is Poisson distributed \cite{27}. Clutter and birth RFSs are assumed to be Poisson RFSs. It is noted that the assumptions made by the PHD filter are strong assumptions for swarming robotics. However, this is a good starting point for an initial proof-of-concept study. 
For a time-step $k$, the PHD recursion for a general intensity function, $v_k(\boldsymbol{\xi})$, from the previous generalized state, $\boldsymbol{\zeta}\in\mathcal{X}$, is given by

\begin{equation}\label{rfs1}
\bar{v}_k(\boldsymbol{\xi})=b(\boldsymbol{\xi})+\int p_s(\boldsymbol{\zeta})f(\boldsymbol{\xi}|\boldsymbol{\zeta})v(\boldsymbol{\zeta})d\boldsymbol{\zeta}+\int \beta(\boldsymbol{\xi}|\boldsymbol{\zeta})v(\boldsymbol{\zeta})d\boldsymbol{\zeta},
\end{equation}
where $b(\boldsymbol{\xi})$, $p_s(\boldsymbol{\zeta})$, and $\beta(\boldsymbol{\xi}|\boldsymbol{\zeta})$  are the agents' birth, survival, and spawn intensity, and $f(\boldsymbol{\xi}|\boldsymbol{\zeta})$ is the target motion model \cite{27}. The bar on $\bar{v}_k(\boldsymbol{\xi})$ denotes that the PHD has been time-updated.
For the measurement update, the equation is given by

\begin{equation} \label{rfs2}
{v_k(\boldsymbol{\xi})=(1-p_d(\boldsymbol{\xi}))\bar{v}_k(\boldsymbol{\xi})}+\sum_{{\bf z}\in Z_k}\frac{p_d(\boldsymbol{\xi})g({\bf z}_k|\boldsymbol{\xi})\bar{v}_k(\boldsymbol{\xi})}{c({\bf z})+\int p_d(\boldsymbol{\zeta})g({\bf z}_k|\boldsymbol{\zeta})\bar{v}_k(\boldsymbol{\zeta})d\boldsymbol{\zeta}},
\end{equation}
where $p_d(\boldsymbol{\xi})$, $g({\bf z}_k|\boldsymbol{\xi})$, and $c({\bf z})$ are the probability of detection, likelihood function, and clutter model of the sensor respectively \cite{27}. By using this recursion, the swarm probabilistic description can be updated. The recursion itself avoids computations that arise from the unknown relation between agents and its measurements, and that the posterior intensity is a function of the generalized state space. Unfortunately, Eqs.~\eqref{rfs1} and \eqref{rfs2} do not contain a closed-form solution and the numerical integration suffers from higher computational time as the state increases due to an increasing number of agents. 
\section{Random Finite Set Control Problem Formulation}
With the introduction of RFS theory from multi-agent tracking applications, a natural extension of RFS theory to the swarm control problem is appealing. We begin the discussion on how the swarm is modelled using the RFS intensity function and present the RFS control problem with the objective based on this model.

The PHD filter recursion given by Eqs. \eqref{rfs1} and \eqref{rfs2}, as mentioned before, can be intractable as the state space increases. Fortunately, a closed-form solution exists if it is assumed that the survival and detection probabilities are state independent (i.e. $p_s(\boldsymbol{\xi})=p_s$ and $p_d(\boldsymbol{\xi})=p_d$), and the intensities of the birth and spawn RFSs are Gaussian mixtures initially presented in Eq. \eqref{gmexample}. 

With the Gaussian mixture assumption, the current and desired intensities are defined as
\begin{equation}\label{funcff}
\bar{\nu}(\boldsymbol{\xi},k)\triangleq\sum_{i=1}^{N_f}w_f^{(i)}\mathcal{N}\left(\boldsymbol{\xi};{\bf m}_f^i,P_f^i\right)=\nu_b(\boldsymbol{\xi},k)+\nu_{p_s}(\boldsymbol{\xi},k)+\nu_\beta(\boldsymbol{\xi},k),
\end{equation}
\begin{equation}\label{funcg}
\nu_{des}(\boldsymbol{\xi},k)\triangleq g(\boldsymbol{\xi})\triangleq\sum_{i=1}^{N_g}w_g^{(i)}\mathcal{N}\left({\boldsymbol{\xi};\bf m}_g^i,P_g^i\right),
\end{equation}
where $w^{(i)}$ are the weights and $\mathcal{N}\left(\boldsymbol{\xi};{\bf m}^i,P^i\right)$ is the probability density function of a $i$th multivariate Gaussian distribution with a mean and covariance corresponding to the peaks and spread of the intensity respectively. The terms $N_f$ and $N_g$ are the total number of multivariate Gaussian distributions in the current and desired intensities, respectively. It is assumed that the desired Gaussian mixture intensity, $\nu_{des}(\boldsymbol{\xi},k)$, is known. Eq. \eqref{funcff} includes the summation of the individual birth ($\nu_b(\boldsymbol{\xi},k)$), spawn ($\nu_\beta$), and survival ($\nu_{p_s}(\boldsymbol{\xi},k)$) Gaussian mixture intensities which simplify to another Gaussian mixture. Note that closed form solutions using Gaussian mixtures exist for cases without the state independent assumption. Additionally, $\sum_{i=1}^{N_f}w_f^{(i)}=N_{\text{total}}(k)$ and $\sum_{i=1}^{N_g}w_g^{(i)}=\bar{N}_{\text{total}}(k)$ where $\bar{N}_{\text{total}}(k)$ is the desired number of agents. The current and desired intensity functions, $\nu(\boldsymbol{\xi},k)$ and $\nu_{des}(\boldsymbol{\xi},k)$, are in terms of the agents' state.
The swarm intensity function can be propagated through updates on the mean and covariance of the Gaussian mixtures as given by
\begin{equation}\label{dynamicswithu}
{\bf m}_{f,k+1}^i=A_k{\bf m}_{f,k}^i+B_k{\bf u}_{f,k}^i,
\end{equation}
\begin{equation}\label{covary}
P_{f,k+1}^i=A_kP_{f,k}^iA_k^T+\Sigma_{\boldsymbol{\epsilon}}.
\end{equation}
The agents' states $\mathbf{x}$ are incorporated in the mean and covariance of the Gaussian mixture intensity. Then given the Gaussian mixture intensities assumption, a control variable is calculated for each component ${\bf u}_{f,k}^i$. \textcolor{black}{Additionally, each Gaussian mixture component may represent many agents since the intensity function integrates to the total number of agents and the number of agents on the field is found using $\sum_{i=1}^{N_f}w_f^{(i)}=N_{\text{total}}(k)$. So control is directly applied to the Gaussian mixture, which may represent single or multiple agents. The weights can be scaled as the number of agents in the swarm change while not effecting the optimal solution found for swarm control. This detail is emphasized in the Results section.} Note that although linear dynamics are used, the dynamics can be modeled as a nonlinear function of the state.

The measurement update is also closed form given by the intensity
\begin{equation}\label{funcf}
{\nu_k(\boldsymbol{\xi},k)=f(\boldsymbol{\xi})=(1-p_d(\boldsymbol{\xi}))\bar{\nu}_k(\boldsymbol{\xi})}+\sum_{{\bf z}\in Z_k}\sum_{j=1}^{N_f}w_k^{(j)}\mathcal{N}\left( \boldsymbol{\xi};{\bf m}_{k|k}^{(j)}({\bf z}),P_{k|k}^{(j)}\right),
\end{equation}
where
\begin{subequations}
\begin{equation}
w_k^{(j)}=\frac{p_d(\boldsymbol{\xi})w_f^{(j)}q^{(j)}({\bf z})}{K({\bf z})+p_d(\boldsymbol{\xi})\sum_{l=1}^{N_f}w_f^{(l)}q^{(l)}({\bf z})},
\end{equation}
\begin{equation}
{\bf m}_{k|k}^{(j)}({\bf z})={\bf m}_{f}^{(j)}+K^{(j)}\left( {\bf z}-H_k{\bf m}_{f}^{(j)}\right),
\end{equation}
\begin{equation}
P_{k|k}^{(j)}=\left( I-K^{(j)}H_k\right)P_f^i,
\end{equation}
\begin{equation}
K^{(j)}=P_f^iH_k^{T}\left( H_kP_f^iH_k^{T}+\Sigma_{\boldsymbol{\sigma}}\right)^{-1},
\end{equation}
\begin{equation}
q_k^{(j)}({\bf z})=\mathcal{N}\left( {\bf z}; H_k{\bf m}_{f}^{(j)},\Sigma_{\boldsymbol{\sigma}}+H_kP_f^iH_k^{T}\right),
\end{equation}
\end{subequations}
which follow closely to the Kalman filter measurement update equations.

Using RFS theory, it is assumed that the individual swarm agents form a Gaussian mixture intensity function in which the means and covariances of the Gaussian mixture are propagated and controlled. An optimal control problem is defined by minimizing the swarm control effort and ``distance" from the desired swarm formation. The objective function for this optimal control problem is defined as:
\begin{equation}\label{distance}
J({\bf u}_1,...,{\bf u}_T)=\sum_{k=1}^T{\bf u}_k^{T}R{\bf u}_k+D(\nu(\boldsymbol{\xi},k),\nu_{des}(\boldsymbol{\xi},k)),
\end{equation}
where $\nu_{des}(\boldsymbol{\xi},k)$ is the desired formation, $R$ is the positive definite control weight matrix, and the ${\bf u}_k$ are the control effort for the Gaussian mixture intensities shown in Eq. \eqref{dynamicswithu}. Both $\nu(\boldsymbol{\xi},k)$ and $\nu_{des}(\boldsymbol{\xi},k)$ are defined over the complete state space which include position and velocity parameters. The distance between Gaussian mixtures, $D(\cdot,\cdot)$, has several closed-form solutions, and it has been used previously to define an objective function for path planning of multi-agent systems \cite{41}.

\textcolor{black}{The key features for the RFS control problem is that it can allow for a unified representation for swarming systems. This unified representation is achieved by minimizing the RFS objective function, Eq. \eqref{distance}, about the swarm intensity statistics given by Eq. \eqref{dynamicswithu} and \eqref{covary}. Thus, it can handle multi-fidelity swarm localization and control between the current and desired spatial distributions. The swarm is treated probabilistically and the bulk motion is modeled which allows the theory to handle large numbers of indistinguishable units with unknown swarm size. This reduces the dimensionality of the state while enabling complex behavior. Naturally, the RFS control problem is formulated to enable complex decision making through RFS theory.}

Two different scenarios can be applied by ``closing-the-loop'' between RFS control and the PHD filter shown in Fig. \ref{closedloop}. Specifically, a single observer can be used to estimate the entire state of the swarm by collecting measurements of each agent in the field. This provides a centralized approach to obtaining estimates and controlling the swarm through RFSs. The other option is to run a local PHD observer on each agent to estimate the state of the swarm. In this case, the observer is limited to an agent's field of view, but it is able to make localized or decentralized control decisions using RFS. For this work, centralized RFS control is explored by using the complete topology obtained from the PHD filter. 
\section{Distributional Distance-Based Cost}
The control objective for the RFS formulation of agents with an unknown distance between the intensities is provided by Eq. \eqref{distance}. The distance metric can be defined using several closed-form solutions for Gaussian mixtures. Then, the corresponding optimal control problem is formulated using several closed-form methods discussed in the next section.
\subsection{Cauchy-Schwarz Divergence}
The Cauchy-Schwarz divergence is based on the Cauchy-Schwarz inequality for inner products of RFS, and it is defined for two RFS with intensities $f$ and $g$ given by 
\begin{equation}\label{cauchy}
D_{CS}(f,g)=-\ln\left(\frac{\left\langle f,g\right\rangle}{\|f\|\|g\|}\right),
\end{equation}
where $\left\langle r(\boldsymbol{\xi}),t(\boldsymbol{\xi}) \right\rangle\triangleq\int r(\boldsymbol{\xi})t(\boldsymbol{\xi})d\boldsymbol{\xi}$ is the $L_2^2$ inner product over generalized the RFS intensities $r(\boldsymbol{\xi})$ and $t(\boldsymbol{\xi})$ \cite{12}. The argument of the logarithm is non-negative because probability densities are non-negative, and it does not exceed one by the Cauchy-Schwarz inequality. The Cauchy-Schwarz divergence can be interpreted as an approximation to the Kullback-Leibler divergence but has a closed-form expression for Gaussian mixtures \cite{12}. This is useful for calculating the distance between two-point processes represented by intensity functions. By substituting the intensities from Eq. \eqref{funcf} and Eq. \eqref{funcg} for $f$ and $g$ respectively, the Cauchy-Schwarz divergence between two Poisson point processes with Gaussian mixture intensities, $D_{CS}(f,g)$, is simplified to
\begin{equation}
\begin{aligned}
D_{CS}(f,g)&=\frac{1}{2}\ln\left( \sum_{j=1}^{N_f}\sum_{i=1}^{N_f}w_f^{(j)}w_f^{(i)}\mathcal{N}({\bf m}_f^j;{\bf m}_f^i,P_f^i+P_f^j)\right)\\
&+\frac{1}{2}\ln\left( \sum_{j=1}^{N_g}\sum_{i=1}^{N_g}w_g^{(j)}w_g^{(i)}\mathcal{N}({\bf m}_g^j;{\bf m}_g^i,P_g^i+P_g^j)\right)\\&-\ln\left( \sum_{j=1}^{N_g}\sum_{i=1}^{N_f}w_g^{(j)}w_f^{(i)}\mathcal{N}({\bf m}_g^j;{\bf m}_f^i,P_g^i+P_f^j)\right).
\end{aligned}
\end{equation}
Note that in the control formulation used, only $\nu(\boldsymbol{\xi},k)$ is assumed to depend on the control ${\bf u}$. Therefore, the term that depends only on $\nu_{des}(\boldsymbol{\xi},k)$ is omitted from the objective function since $\nu_{des}(\boldsymbol{\xi},k)$ does not depend on ${\bf u}$. 

\begin{figure}[h]
\begin{centering}
    \subfigure[Cauchy-Schwarz Divergence]{
      \includegraphics[keepaspectratio,trim={.35cm .05cm .25cm .3cm},clip,width=.48\textwidth]{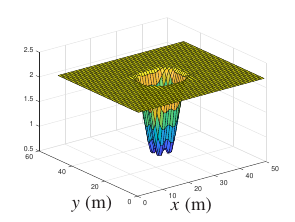}
      \label{fig:CS}}
          \subfigure[$L_2^2$ Distance]{
    \includegraphics[keepaspectratio,trim={.35cm .05cm .25cm .3cm},clip,width=.48\textwidth]{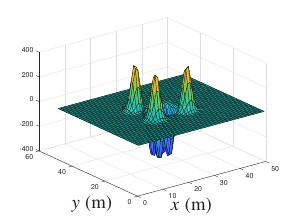}
    \label{fig:L2}}  
              \subfigure[$L_2^2$ + Quadratic Term]{
      \includegraphics[keepaspectratio,trim={.35cm .05cm .25cm .3cm},clip,width=.48\textwidth]{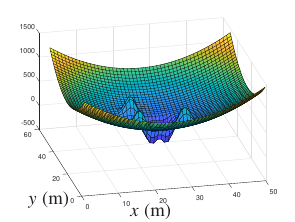}
      \label{fig:L2q}} 
        \caption{ (a), (b), and (c) are the surface plots with the corresponding distributional distance-based costs. The current and desired intensity are initialized at ($\pm 3$,$\pm 3$) and ($\pm 1$,$\pm 1$), respectively.}\label{contours}
\end{centering}
\end{figure}

Figure \ref{fig:CS} shows the surface plot using the Cauchy-Schwarz divergence for four Gaussian mixtures in the swarm at an initial time instance which designates the distributional distance-based cost of the objective function. The four Gaussian mixtures start with initial conditions of ($\pm 3$,$\pm 3$) in a square grid. The desired intensity is set as ($\pm 1$,$\pm 1$) in a square grid. From the surface plot, each initial intensity has hills while the desired intensity has valleys. The goal is to minimize the objective function, thus, an optimization method (e.g. the Quasi-Newton method ) determines a control solution which minimizes the objective. Since the desired intensity in Fig. \ref{fig:CS} is located at a minimum in the objective surface plot, the optimization method finds a control input to move towards that point. The opposite occurs with the hills (current intensity). The minimization finds a control solution that moves away from the hills, and thus gives individual current Gaussian mixtures collision avoidance attributes. Therefore in the minimization of the objective function, each Gaussian mixture will repel each other while moving towards the desired Gaussian mixtures through time. Although the Cauchy-Schwarz divergence has a repelling effect, collision avoidance is not guaranteed, but the distance does encourage collision-reducing trajectory solutions. If the initial intensity is too large compared to the desired intensity, it will take longer for the four Gaussian mixtures to converge to the desired values or diverge due to the optimization getting stuck in local minima (the flat plane). Also, the repelling effect due to the hills are relatively small. Thus, the Cauchy-Schwarz divergence may not be the fastest converging solution for the objective function minimization.
\subsection{$L_2^2$ Distance}
Alternatively, the distance between two Poisson point processes with Gaussian mixture intensities can be determined by using the $L_2^2$ distance between the intensities. The $L_2^2$ distance is given by
\begin{equation}\label{l2gen}
D_{L_2^2}(f,g)=\int\left( f-g\right)^2d\boldsymbol{\xi}=||f-g||^2,
\end{equation}
where the close-form solution for Gaussian mixture intensities is simplified to
\begin{equation}\label{l2eqn}
\begin{aligned}
D_{L_2^2}(f,g)&=\sum_{j=1}^{N_f}\sum_{i=1}^{N_f}w_f^{(j)}w_f^{(i)}\mathcal{N}\left({\bf m}_f^j;{\bf m}_f^i,P_f^i+P_f^j\right)\\&+\sum_{j=1}^{N_g}\sum_{i=1}^{N_g}w_g^{(j)}w_g^{(i)}\mathcal{N}\left({\bf m}_g^j;{\bf m}_g^i,P_g^i+P_g^j\right)\\&-2\sum_{j=1}^{N_g}\sum_{i=1}^{N_f}w_g^{(j)}w_f^{(i)}\mathcal{N}\left({\bf m}_g^j;{\bf m}_f^i,P_g^i+P_f^j\right).
\end{aligned}
\end{equation}
The $L_2^2$ distance is stationary, i.e. gradients are zero, when intensities $f$ and $g$ are equal. That is, the cost is minimum when the target $g$ is reached from any intensity $f$. 

The $L_2^2$ distance follows the property of the Bregman divergence which has an additional property of convexity \cite{36}. The distance, given by 
\begin{equation}\label{breg}
D_F\left( f,g \right) =F(f)-F(g)-\langle\nabla F(g),f-g\rangle,
\end{equation}
is convex if $F(\cdot)$ is strictly convex and continuously differentiable on a closed convex set \cite{36}. A list of strictly convex functions are listed in \cite{36}. For this work, the squared Euclidean  distance $F(f)=f^2$ was used to generate the Bregman divergence given by
\begin{equation}
D_F\left( f,g \right) =\left\langle f,f \right\rangle + \left\langle g,g \right\rangle-2\left\langle f,g \right\rangle,
\end{equation}
which is in the same exact form of Eq. \eqref{l2eqn}.
Figure \ref{fig:L2} shows the surface plot using the $L_2^2$ distance for a 4 Gaussian mixture swarm for the same example as the Cauchy-Schwarz divergence. The initial intensity has more defined hills compared to the Cauchy-Schwarz divergence. Thus, the initial Gaussian mixtures have a stronger repelling effect upon one another. Also, the desired Gaussian mixtures have large valleys that create a large attraction effect for each initial Gaussian mixture to move to. Thus, the optimization solution will be faster in the $L_2^2$ distance case. Unfortunately, the $L_2^2$ distance suffers from a similar issue to the Cauchy-Schwarz divergence. If the initial conditions increase farther away from the desired intensity, the optimization may take much longer or get stuck in local minima due to a flat surface away from the desired intensity.

\subsection{$L_2^2$ Distance with Quadratic Term}
The issue of convergence remains for the $L_2^2$ distance when the initial states are farther away from the desired intensity. To achieve faster convergence, an additional term is added to the $L_2^2$ distance to shape the gradient descent through a quadratic term as given by 
\begin{equation}\label{modifyL2}
{D_{L_2^2mod}(f,g)=D_{L_2^2}(f,g)}-\alpha{ \sum_{j=1}^{N_g}\sum_{i=1}^{N_f}w_g^{(j)}w_f^{(i)}\ln\left(\mathcal{N}({\bf m}_g^j;{\bf m}_f^i,P_g^i+P_f^j)\right)},
\end{equation}
where $\alpha$ is a fixed or changing parameter. Unfortunately, adding the quadratic term to the $L_2^2$ distance does not make the objective function stationary at $f=g$. To alleviate this issue, the $\alpha$ parameter is included with the quadratic term to relax the contribution of the gradient to the $L_2^2$ stationary point.
By substituting Eq. \eqref{l2eqn} into Eq. \eqref{modifyL2}, the equation becomes
\begin{equation}\label{modifiedL2}
\begin{aligned}
D_{L_2^2mod}(f,g)&=\sum_{j=1}^{N_f}\sum_{i=1}^{N_f}w_f^{(j)}w_f^{(i)}\mathcal{N}({\bf m}_f^j;{\bf m}_f^i,P_f^i+P_f^j)\\&+\sum_{j=1}^{N_g}\sum_{i=1}^{N_g}w_g^{(j)}w_g^{(i)}\mathcal{N}({\bf m}_g^j;{\bf m}_g^i,P_g^i+P_g^j)\\&-2\sum_{j=1}^{N_g}\sum_{i=1}^{N_f}w_g^{(j)}w_f^{(i)}\mathcal{N}({\bf m}_g^j;{\bf m}_f^i,P_g^i+P_f^j)\\&- \alpha\sum_{j=1}^{N_g}\sum_{i=1}^{N_f}w_g^{(j)}w_f^{(i)}\ln\left(\mathcal{N}({\bf m}_g^j;{\bf m}_f^i,P_g^i+P_f^j)\right).
\end{aligned}
\end{equation}
Note that this term is referred as quadratic, although it may be more appropriate to call it quadratic-like.
Figure \ref{fig:L2q} shows the surface plot using Eq. \eqref{modifiedL2} for the same 4 Gaussian mixture swarm used in the Cauchy-Schwarz divergence. Compared to the $L_2^2$ distance, the initial and desired intensities provide the hills and valleys necessary to obtain convergence. However, as the initial intensity move outwards, the surface map decreases in a quadratic fashion instead of staying flat. This prevents the optimization from converging to a local minima. Instead, the additional quadratic term allows convergence to the desired intensity (global minima). Thus, the optimization can occur at any point to reach convergence.

Traditional LQR based solutions are not applicable to the minimization of the objective function, Eq. \eqref{modifiedL2}, since the $L_2^2$ terms are nonquadratic \cite{11}. 
The minimization of the objective function in discrete time is
\begin{equation}\label{overalleqn}
\begin{aligned}
\min_{{\bf u}_k,k=1,...,T}J({\bf u}_1,...,{\bf u}_T)&=\sum_{k=1}^T\ {\bf u}_k^{T}R{\bf u}_k+\sum_{j=1}^{N_f}\sum_{i=1}^{N_f}w_{f,k}^{(j)}w_{f,k}^{(i)}\mathcal{N}({\bf m}_{f,k}^j;{\bf m}_{f,k}^i,P_{f,k}^i+P_{f,k}^j)\\&+\sum_{j=1}^{N_g}\sum_{i=1}^{N_g}w_{g,k}^{(j)}w_{g,k}^{(i)}\mathcal{N}({\bf m}_{g,k}^j;{\bf m}_{g,k}^i,P_{g,k}^i+P_{g,k}^j)\\&-2\sum_{j=1}^{N_g}\sum_{i=1}^{N_f}w_{g,k}^{(j)}w_{f,k}^{(i)}\mathcal{N}({\bf m}_{g,k}^j;{\bf m}_{f,k}^i,P_{g,k}^i+P_{f,k}^j)\\&- \alpha\sum_{j=1}^{N_g}\sum_{i=1}^{N_f}w_{g,k}^{(j)}w_{f,k}^{(i)}\ln\left(\mathcal{N}({\bf m}_{g,k}^j;{\bf m}_{f,k}^i,P_{g,k}^i+P_{f,k}^j)\right),
\end{aligned}
\end{equation}
\begin{equation}\label{qwer}
\begin{split}
\text{Subject} \hspace{5pt}\text{to}:{\bf m}_{f,k+1}^i=A_k{\bf m}_{f,k}^i+B_k{\bf u}_{f,k}^i,\\
P_{f,k+1}^i=A_kP_{f,k}^iA_k^T+\Sigma_{\boldsymbol{\epsilon}},
\end{split}
\end{equation}
where ${\bf u}_k=[({\bf u}_{f,k}^1)^T,\cdots,({\bf u}_{f,k}^{N_f})^T]^T$ is the collection of all control variables. 
Therefore, control solutions are found by either using DDP where the objective function is quadratized by taking a Taylor series approximation about a nominal trajectory or using optimization techniques (e.g. the Quasi-Newton method) where the nonquadratic objective function is used directly to find an optimal control solution.
\textcolor{black}{The number of agents is not a controlled quantity and can be normalized out of the objective function (this can be shown by examining Eq.~\eqref{overalleqn}). In this work we do not desire to control the number of agents but rather control individual agents to track the desired ``concentration” or distribution of agents. Under the proposed formulation, the distribution of agents is indeed a controllable quantity using the control inputs defined in Eq.~\eqref{overalleqn}.}

To obtain efficient control solutions for large time horizons, MPC is used directly with the control techniques. Quasi-Newton handles the nonlinearities in the objective function, and it provides an initial basis in comparing the time-history responses for RFS control using different distributional distance-based costs. Next, RFS control is extended to MPC with DDP which approximates (quadratizes) the objective function for value iteration to provide quick and reliable convergence to locally-optimal control solutions. DDP and MPC are discussed in the Appendix. The RFS control solution is demonstrated on spacecraft swarm relative motion simulation with and without perfect information combining the GM-PHD filter and DDP (formed as ILQR) in a closed-loop fashion given in Fig. \ref{closedloop}. In a single loop, the RFS control is determined from optimizing the objective containing the distributional distance based cost between the estimate and the desired intensity, the swarm dynamics are updated with new spawn, birth, or death of agents in the field, and measurements including clutter are incorporated into the overall system before a GM-PHD filter estimate is determined for control again. From this RFS-based architecture, the ability to determine an estimate of the cardinality and states of the swarm which is used directly for control using ILQR is realized.

As previously discussed for this work, the topology underlying the RFS control is complete and uses the complete graph in a centralized manner. Thus, the computational load for control of the swarm is centralized. It is computationally feasible to perform centralized control from a single agent, although a separate ground station may be necessary to perform difficult control computations.

\section{Dynamical Models}\label{dynamicmodelsec}
To show viability of optimal swarm control via RFS, an acceleration model and a relative motion model, both linear systems, are used to describe rover and satellite dynamics, respectively. The dynamic equations of individual agents are used here to describe the dynamics of the Gaussian mixture components (means) given by the control objective Eqs. \eqref{overalleqn} and \eqref{qwer}. Since linear dynamics are used, the DDP term can be expressed as ILQR.
\subsection{Acceleration Model}
On a 2D plane, the linear time-invariant (LTI) system of each agent can be described by the continuous state and control matrices
\begin{equation} \label{qMatrix}
 A_c = \begin{bmatrix}0& 0 & 1 &0\\ 0 & 0&0&1\\0&0&0&0\\0&0&0&0 \end{bmatrix}, \hspace{12pt} B_c = \begin{bmatrix} 0&0 \\0&0\\1&0\\0&1\\  \end{bmatrix},
\end{equation}
and a state vector $\mathbf{x}=\left[x,y,\dot{x},\dot{y} \right]^{T}$. Both $x$ and $y$ are defined to be the 2D positions of the agent respectively.
The $A_c$ and $B_c$ matrices are discretized along a fixed time interval utilizing a zero-order hold assumption for the control (i.e. control is held constant over the time-interval). This results in discretized $A$ and $B$ matrices for the state-space equation,
\begin{equation}\label{discretization}
{\bf x}_{k+1}=A{\bf x}_{k}+B{\bf u}_{k}.
\end{equation}
 
\subsection{Relative Motion using Clohessy-Wiltshire Equations}
For a spacecraft in low Earth orbit, the relative dynamics of each spacecraft (agent), to a chief spacecraft in circular orbit, is given by the Clohessy-Wiltshire equations \cite{42}
\begin{subequations}
\begin{equation}
\ddot{x}=3n^2x+2n\dot{y}+a_x,
\end{equation}
\begin{equation}
\ddot{y}=-2n\dot{x}+a_y,
\end{equation}
\begin{equation}
\ddot{z}=-n^2z+a_z,
\end{equation}
\end{subequations}
where $x$, $y$, and $z$ are the relative positions in the orbital local-vertical local-horizontal (LVLH) frame and $a_x$, $a_y$, and $a_z$ are the accelerations in each axis respectively. The variable $n$ is defined as the orbital frequency given by
\begin{equation}
n=\sqrt{\frac{\mu}{a^3}},
\end{equation}
where $\mu$ is the standard gravitational parameter and $a$ is the radius of the circular orbit. The continuous state-space representation is given by
\begin{equation} \label{qqMatrix}
 A_c = \begin{bmatrix}0&0& 0 & 1 &0&0\\ 0 & 0&0&0&1&0\\0&0&0&0&0&1\\3n^2&0&0&0&2n&0\\0&0&0&-2n&0&0\\0&0&-n^2&0&0&0 \end{bmatrix}, \hspace{12pt} B_c = \begin{bmatrix} 0&0&0 \\0&0&0\\0&0&0\\1&0&0\\0&1&0\\0&0&1  \end{bmatrix},
\end{equation}
with a state vector $\mathbf{x}=\left[ x, y, z, \dot{x}, \dot{y}, \dot{z}\right]$ and a control input $\mathbf{u}=\left[ a_x, a_y, a_z\right]^{T}$. These equations are discretized similarly to the acceleration model discussed previously.

\section{Results}
The first goal is to generalize RFSs for control of large collaborative swarms to form and test behaviors of several different RFS-based distance measures for control. Using the acceleration model, which is discretized from Eq. \eqref{qMatrix} to Eq. \eqref{discretization}, a 4 Gaussian mixture swarm on a 2-D plane is initialized in a square grid where the mixtures 1, 2, 3, and 4 are defined counterclockwise starting on the first quadrant. With the 4 Gaussian mixture swarm, three different test cases are implemented to bring the intensity to the target trajectories and to test the distributional distance-based costs and control theory involved from the RFS formulation. The first test case compares the $L_2^2$ distance with varying initial conditions in a square grid with the $L_2^2$ plus quadratic distance with four desired Gaussian mixtures located at ($\pm$1,$\pm$1) using Quasi-Newton MPC. An $L_2^2$ plus quadratic distance comparison is also done using ILQR. The last two cases present the Quasi-Newton MPC and ILQR control using the $L_2^2$ distance with a quadratic term and varying desired Gaussian mixtures. For Case 2, three target destinations are located at ($\pm$1,1) and (-1,-1). Lastly, in Case 3, five target destinations are located at ($\pm$1,$\pm$1) and (0,0). 

The second goal is to form and apply multi-agent estimation and control of large collaborative swarms in the presence of unknown number of agents, clutter, and noise using RFS theory. This is applied directly to the satellite relative motion problem using the Clohessy-Wiltshire Equations. Specifically, the $L_2^2$ plus quadratic distance is used for spacecraft formation flight. A 77 Gaussian mixture swarm is initialized uniformly random between -1 and 1 on a 2D plane. Assuming that the spacecraft swarm is at lower Earth orbit, the goal for the spacecraft is to track a rotating star pattern moving counterclockwise at an orbital frequency of $n$.

\textcolor{black}{For these two separate examples, it should be emphasized that each Gaussian mixture component may represent many agents since the intensity function integrates to the total number of agents. For the initial example, many agents are represented by each Gaussian mixture, and the number of agents on the field must be calculated using $\sum_{i=1}^{N_f}w_f^{(i)}=N_{\text{total}}(k)$. So individual agents follow the control law that is applied to their specific Gaussian mixture. As mentioned before, the weights of the desired RFS do not affect the optimal solution and therefore, for simplicity, this initial goal uses weights of $w_k=1$ for all $k$. For the second example which incorporates the Clohessy-Wiltshire relative motion model, the perfect and imperfect information scenarios assume the number of agents are equal to the number of Gaussian mixtures (i.e. $w_f^{(i)}=1$ for all $k$) and the number of agents are estimated from the GM-PHD filter at each time-step $k$, respectively. Thus, the individual agents are controlled directly.}

\subsection{Acceleration Model}
\subsubsection{Case 1: $L_2^2$ vs. $L_2^2$+Quadratic Term, Four Desired Gaussian Mixtures }
\begin{figure}[!ht]
\begin{centering}
    \subfigure[$L_2^2$ Distance Trajectory]{
\includegraphics[keepaspectratio,trim={.5cm .15cm .05cm .23cm},clip, width=0.48\textwidth]{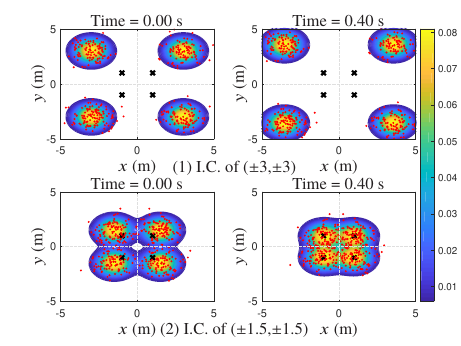}
\label{Case0}}
\subfigure[$L_2^2$ and $L_2^2$+Quadratic Distance Position Time History]{
\includegraphics[keepaspectratio,trim={.5cm .15cm .05cm .23cm},clip,width=0.48\textwidth]{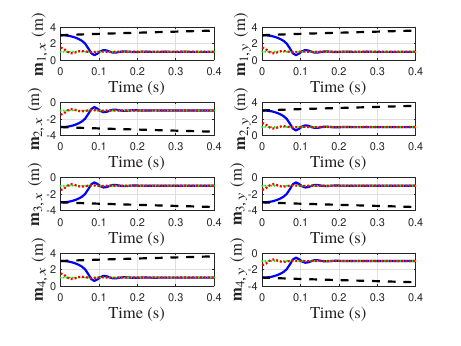}
\label{fig:Case01}}  
              \subfigure[$L_2^2$+Quadratic Distance Trajectory]{
\includegraphics[keepaspectratio,trim={.5cm .15cm .05cm .23cm},clip, width=0.48\textwidth]{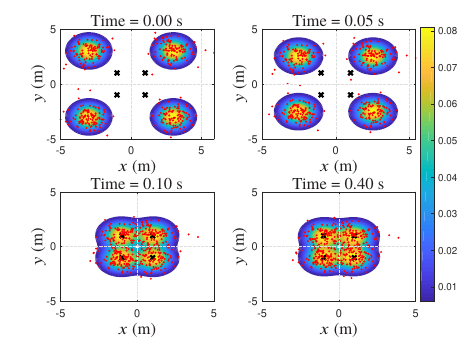}
\label{fig:case11}} 
        \caption{Case 1: (a) and (c) show the controlled trajectories using the acceleration model and Quasi-Newton MPC. (b) shows the intensity mean responses from the trajectories.}\label{fig:l2l2q}
\end{centering}
\end{figure}

For Case 1, four swarm Gaussian mixtures are controlled to move towards the desired intensity at initial conditions farther away (square grid at ($\pm$3,$\pm$3)) and closer to (square grid at ($\pm$1.5,$\pm$1.5)) the desired intensity as shown by mean responses given by the black-dashed and red-dotted lines in Fig. \ref{fig:Case01} respectively. From the trajectory snapshots given by Figure 2(a1), initial conditions that are far from the desired intensity do not have a converging control solution. From the surface visualization in Fig. \ref{fig:L2}, the general plane is flat in areas away from the desired targets and the current agents' states. Therefore, optimization using Quasi-Newton MPC is more difficult in these flat areas and may not converge to a solution. If the current intensity is initialized much closer to the desired intensity as shown in Figure 2(a2), the flatness in the general plane is minimal, and the optimization step in Quasi-Newton MPC converges to a solution. By using the $L_2^2$ distance, converging control solutions can only be found for initial conditions and target destinations that are close.
For the $L_2^2$ plus quadratic distance, four swarm Gaussian mixtures move towards the four desired Gaussian mixtures given by the blue-solid lines (mean responses) in Fig. \ref{fig:Case01}. 
Figure \ref{fig:case11} shows the trajectory snapshots and final states of each of the swarm Gaussian mixtures during the simulation. The target destinations are plotted as black x's. The red dots are the individual swarm agents that form the Gaussian mixture intensities. From the figure, all four mixtures converge to the desired mixtures in approximately 0.17 seconds and approximately 0.03 of steady-state error between the mixtures' position to the desired intensity. In comparison to only the $L_2^2$ distance, Fig. \ref{fig:Case01} shows that for small distances between the initial state and the desired intensity, the $L_2^2$ distance is sufficient for state convergence, but as the distance increases, the $L_2^2$ distance diverges away. By adding the quadratic term to $L_2^2$, the optimization step can directly determine the minimum for the control solution shown in Fig. \ref{fig:L2q}. Therefore, the desired intensity attracts the current swarm intensity at distances that fail for only $L_2^2$ distance given by Fig. \ref{fig:Case01}.

\begin{figure}[!ht]
\begin{centering}
    \subfigure[$L_2^2$+ Quadratic Distance Trajectory]{
\includegraphics[keepaspectratio,trim={.5cm .15cm .05cm .23cm},clip, width=0.48\textwidth]{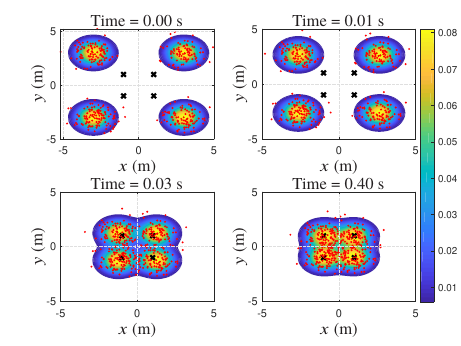}
\label{fig:Case1ilqr}}
\subfigure[$L_2^2$+Quadratic Distance Position Time History]{
\includegraphics[keepaspectratio,trim={.5cm .15cm .05cm .23cm},clip, width=0.48\textwidth]{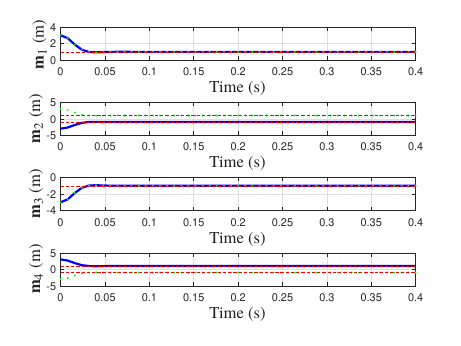}
\label{fig:Case01ilqr}}  
        \caption{Case 1: 4 Gaussian mixture swarm controlled to four desired Gaussian mixtures via ILQR. (a) shows the trajectories for the swarm and (b) shows the position time history.} \label{fig:l2l2q2}
\end{centering}
\end{figure}

The $L_2^2$ plus quadratic distance is also extended to ILQR. Figure \ref{fig:Case1ilqr} shows the trajectory snapshots and final states of the simulation. All four Gaussian mixtures converge to the desired intensity in approximately 0.03 seconds and approximately 0.01 of steady-state error as shown in Fig. \ref{fig:Case01ilqr}. In this figure, the x responses, y responses, and the desired intensity are given by blue, green, and red lines respectively.  The entire simulation horizon is used to provide the prediction horizon for the ILQR trajectory. Even with a quadratic approximation of the objective function, ILQR is able to find control solutions that follow the $L_2^2$ plus quadratic characteristics that are presented using Quasi-Newton MPC. 
\subsubsection{Case 2: Three Desired Gaussian Mixtures}
Case 2 illustrates the effect of three desired Gaussian mixtures on the final trajectories of the four swarm Gaussian mixtures using Quasi-Newton MPC and ILQR.
\begin{figure}[!htb]
\begin{centering}
    \subfigure[$L_2^2$+ Quadratic Distance Trajectory]{
\includegraphics[keepaspectratio,trim={.5cm .15cm .05cm .23cm},clip, width=0.48\textwidth]{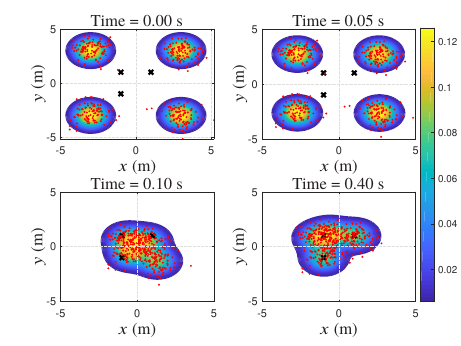}
\label{fig:Case2sub2}}
\subfigure[$L_2^2$+Quadratic Distance Position Time History]{
\includegraphics[keepaspectratio,trim={.5cm .15cm .05cm .23cm},clip, width=0.48\textwidth]{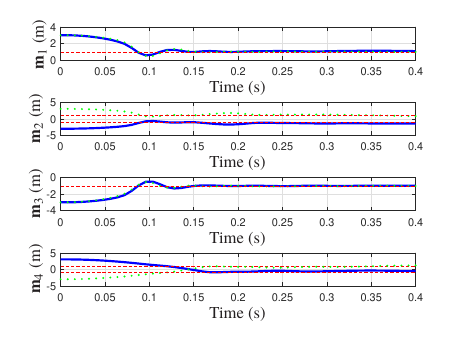}
\label{fig:Case2sub1}}  
        \caption{Case 2: 4 Gaussian mixture swarm controlled to three desired Gaussian mixtures via Quasi-Newton MPC. (a) shows the trajectories for the swarm and (b) shows the position time history.} \label{fig:l2l2q3}
\end{centering}
\end{figure}
Using Quasi-Newton MPC, the current swarm intensity converges as given by the position time-history in Fig. \ref{fig:Case2sub1}. The trajectories for mixture 1 and mixture 3 reach their target, but mixtures 2 and 4 reach the third target with approximately 0.42 and 0.50 of steady-state error with 0.20 and 0.16 seconds of settling time respectively. From Fig. \ref{fig:Case2sub2}, it can be visually shown where the swarm intensity is located relative to the desired intensity at each time-step. The results obtained follow directly from the RFS control theory using the $L_2^2$ plus quadratic distance term. By using this $L_2^2$ with a quadratic term in the objective function, the current intensity will attract towards the desired intensity while repulsing away from each other. This can be seen in the surface map shown in Fig. \ref{fig:L2q}, where the hills are areas of repulsion and valleys, are areas of attraction. Thus, for Quasi-Newton MPC, mixtures 2 and 4 are attracted to the same target, but they stay away from each other. 
\begin{figure}[!htb]
\begin{centering}
    \subfigure[$L_2^2$+ Quadratic Distance Trajectory]{
\includegraphics[keepaspectratio,trim={.5cm .15cm .05cm .23cm},clip, width=0.48\textwidth]{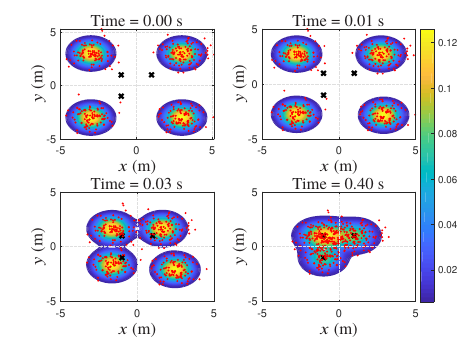}
\label{fig:Case2ilqr}}
\subfigure[$L_2^2$+Quadratic Distance Position Time History]{
\includegraphics[keepaspectratio,trim={.5cm .15cm .05cm .23cm},clip, width=0.48\textwidth]{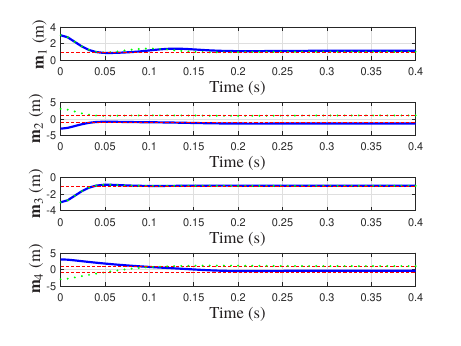}
\label{fig:Case02ilqr}}  
        \caption{Case 2: 4 Gausssian mixture swarm controlled to three desired Gaussian mixtures via ILQR. (a) shows the trajectories for the swarm and (b) shows the position time history.} \label{fig:l2l2q4}
\end{centering}
\end{figure}
This case is also extended to ILQR. Figures \ref{fig:Case2ilqr} and \ref{fig:Case02ilqr} show the trajectory snapshots and time-history of the same swarm using ILQR. As discussed previously, due to the approximation of the objective function, the mixtures 2 and 4 converged in 0.03 and 0.15 seconds with approximately 0.01 and 0.42 of steady-state error. 
By comparing the time-histories in Fig. \ref{fig:Case02ilqr} and \ref{fig:Case2sub1}, the fourth intensity using ILQR follows very similarly to the MPC method. Therefore, there is a degree of accuracy in the approximation of the objective function to minimize for ILQR that allows the attraction of individual mixtures to the desired intensity while repulsing away from each other. 

\subsubsection{Case 3: Five Desired Gaussian Mixtures}
\begin{figure}[!htb]
\begin{centering}
    \subfigure[$L_2^2$+ Quadratic Distance Trajectory]{
\includegraphics[keepaspectratio,trim={.5cm .15cm .05cm .23cm},clip, width=0.48\textwidth]{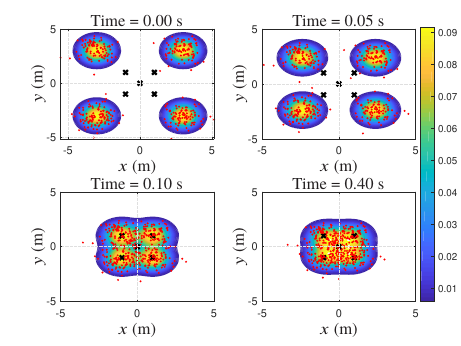}
\label{fig:Case3sub2}}
\subfigure[$L_2^2$+Quadratic Distance Position Time History]{
\includegraphics[keepaspectratio,trim={.5cm .15cm .05cm .23cm},clip, width=0.48\textwidth]{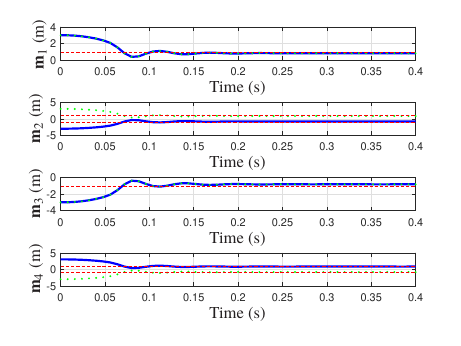}
\label{fig:Case3sub1}}  
        \caption{Case 3: 4 Gaussian mixture swarm controlled to five desired Gaussian mixtures via Quasi-Newton MPC. (a) shows the trajectories for the swarm and (b) shows the position time history.} \label{fig:case35des}
\end{centering}
\end{figure}
Case 3 shows the effect of five desired Gaussian mixtures with the four swarm Gaussian mixtures using Quasi-Newton MPC and ILQR.
Figure \ref{fig:Case3sub1} shows the time histories for all the mixtures using Quasi-Newton MPC.
The trajectory snapshots of the Gaussian mixtures are visually shown relative to the desired Gaussian mixtures in Fig. \ref{fig:Case3sub2}. From Fig. \ref{fig:Case3sub1}, the intensity converges in 0.19 seconds with a steady state error of approximately 0.17 which follow the theory as expected. Since the swarm Gaussian mixtures are far from each other, the effects of repulsion are minimal. Also, the mixtures are attracted to the four desired Gaussian mixtures that make up a square, but they are also attracted to the desired Gaussian mixture at the origin. This is due to the minimization of the objective function that has both an $L_2^2$ and a quadratic term where the individual mixtures will attract towards the desired intensity. Since there is an additional desired Gaussian mixture at the origin, all four swarm Gaussian mixtures are affected by the origin as they are moving towards the 4 square desired Gaussian mixtures. Thus, compared to Case 1 with only four desired Gaussian mixtures, the swarm intensity, in this case, will have a steady-state error due to the attraction to the additional desired Gaussian mixture.
\begin{figure}[!htb]
\begin{centering}
    \subfigure[$L_2^2$+ Quadratic Distance Trajectory]{
\includegraphics[keepaspectratio,trim={.5cm .15cm .05cm .23cm},clip, width=0.48\textwidth]{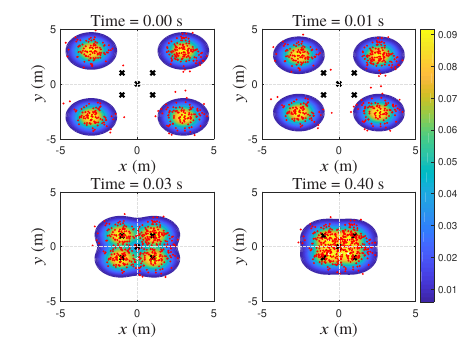}
\label{fig:Case3ilqr}}
\subfigure[$L_2^2$+Quadratic Distance Position Time History]{
\includegraphics[keepaspectratio,trim={.5cm .15cm .05cm .23cm},clip, width=0.48\textwidth]{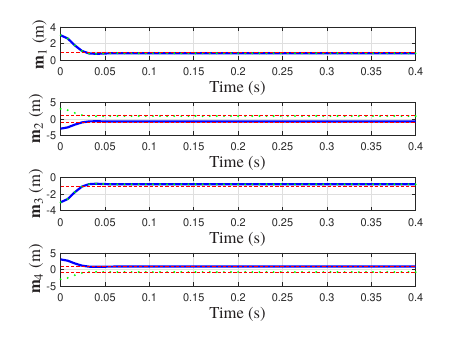}
\label{fig:Case03ilqr}}  
        \caption{Case 3: 4 Gaussian mixture swarm controlled to five desired Gaussian mixtures via ILQR. (a) shows the trajectories for the swarm and (b) shows the position time history.} \label{fig:case3desilqr}
\end{centering}
\end{figure}
ILQR is also used to show how five desired Gaussian mixtures affect the quadratization of the $L_2^2$ plus quadratic objective function. Figures \ref{fig:Case3ilqr} and \ref{fig:Case03ilqr} show the trajectory snapshots and time-history respectively. The swarm converges in 0.03 seconds and 0.12 of steady-state error. This steady-state error shows the attraction of the desired Gaussian mixture at the origin which follows directly from results from the $L_2^2$ plus quadratic distance given by Fig. \ref{fig:L2q}. 
\subsection{Clohessy-Wiltshire Relative Motion}
\subsubsection{Relative Motion with Perfect Information}
\begin{figure}[!htb]
\begin{centering}
    \subfigure[Clohessy-Wiltshire Trajectory Snapshots]{
\includegraphics[keepaspectratio,trim={.40cm .15cm .05cm .23cm},clip, width=0.48\textwidth]{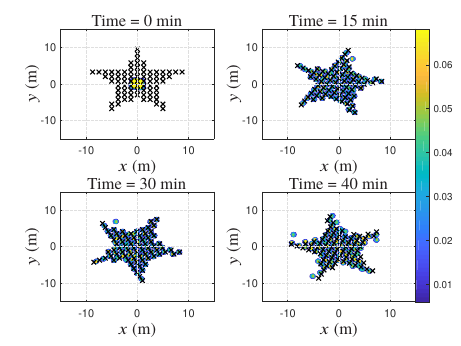}
\label{fig:cwh1}}
\subfigure[Control Input]{
\includegraphics[keepaspectratio,trim={.40cm .15cm .05cm .23cm},clip, width=0.48\textwidth]{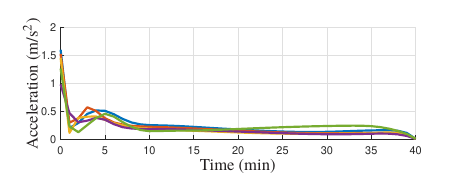}
\label{fig:cwh2}}  
        \caption{ 77 Gaussian mixture spacecraft swarm controlled to a rotating star target via ILQR with perfect information. (a) shows the trajectories and (b) shows the acceleration for five spacecraft intensities.}
\end{centering}
\end{figure}

For the spacecraft relative motion, 77 Gaussian mixtures are birthed at the initial time from uniformly random initial conditions between -1 and 1 m from the chief satellite in a circular orbit. This is similar to the setup in \cite{eren2018density} and follows Fig. \ref{closedloop} without the Swarm Estimation block since it is assumed the state information received throughout the simulation is perfect. Additionally, no agents dies or spawn during the simulation. The goal is to control the spacecraft into a moving star-shaped pattern. Both the spacecraft and the rotating star pattern have an orbital frequency of $n=0.00110678$ rad/s. Figure \ref{fig:cwh1} shows the trajectory snapshots of the spacecraft (contours) and the desire Gaussian mixtures (black x's) using ILQR and the $L_2^2$ plus quadratic distance. The Gaussian mixtures, represented by each contour, can be safely assumed to contain a single agent. As time progresses, agents converge quickly into the formation and maintain the formation for the simulation time of 40 min. A few agents lag behind the desired targets due to the repelling effect from their proximity to other agents in the swarm. Figure \ref{fig:cwh2} shows the acceleration for five agents to maintain the star formation. From these results, control using RFS can be expanded to physical spacecraft systems and can be used for moving targets.
\subsubsection{Relative Motion with Imperfect Information}
\begin{figure}[!htb]
\begin{centering}
    \subfigure[Cardinality]{
\includegraphics[keepaspectratio,trim={.40cm .15cm .05cm .23cm},clip, width=0.48\textwidth]{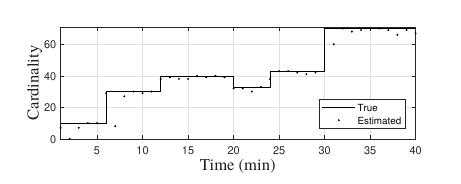}
\label{cardinality}}
\subfigure[Time History]{
\includegraphics[keepaspectratio,trim={.40cm .15cm .05cm .23cm},clip, width=0.48\textwidth]{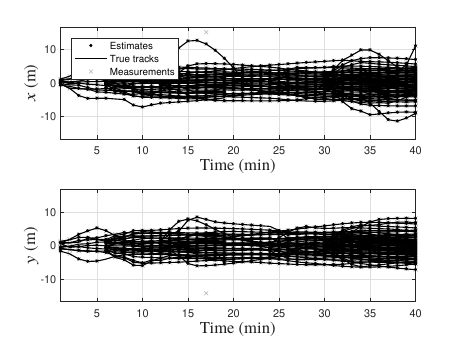}
\label{timehis}}  
\subfigure[Clohessy-Wiltshire Trajectory Snapshots]{
\includegraphics[keepaspectratio,trim={.40cm .15cm .05cm .23cm},clip, width=.95\textwidth]{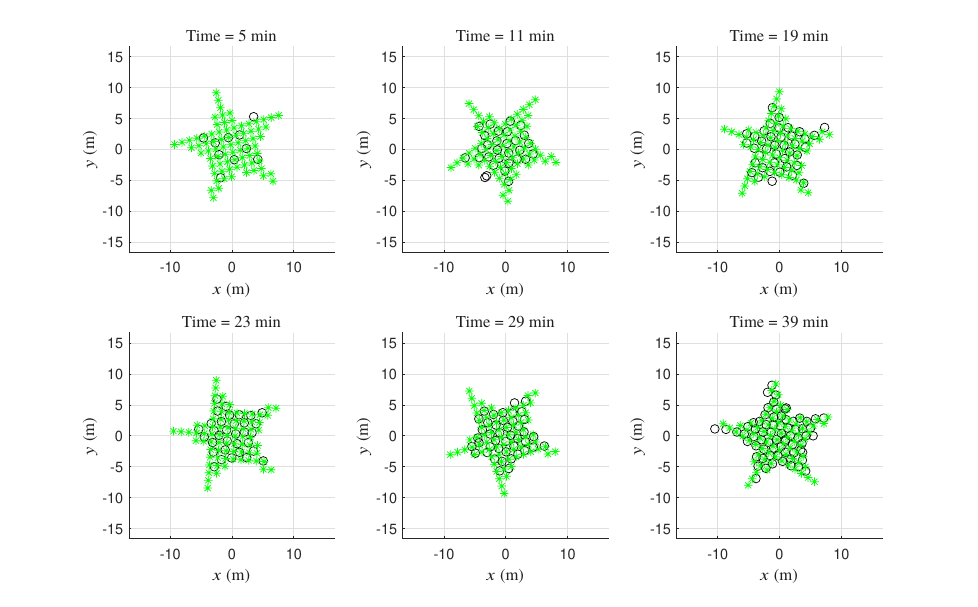}
\label{snapshot}}

        \caption{(c) shows the 77-agent spacecraft swarm controlled via ILQR to a rotating star target with imperfect information. (a) and (b) plot cardinality and the tracks, respectively.}
\end{centering}
\end{figure}

Next, the imperfect information (i.e. process, measurement, and clutter noise) is included in the simulation. In order to control with imperfect information, the GM-PHD filter is used in the Swarm Estimation block in Fig. \ref{closedloop} with the RFS control method. The GM-PHD filter determines the estimates of the intensities which is used for RFS control. The problem was altered to be more complicated by including differing birth and death times for the agents. With the addition of imperfect information and the added complication of changing number of agents, using the GM-PHD filter provides accurate estimates of the agents through time which allows for RFS control in the loop. Figure \ref{cardinality} shows the cardinality or number of agents in the swarm through time. The solid line is the true number of agents while the dotted line shows the estimate at each time-step. At each time-step, the agent estimates are fed through the RFS control using ILQR to obtain a control input for each agent. Then, the estimates are controlled and fed back to the GM-PHD filter at the next time-step. Figure \ref{snapshot} shows the snapshots of the controlled agents (black circles) and targets (green stars) at each time-step. Figure \ref{timehis} shows the time history for the true agents (solid lines), estimated agents (black dots), and overall measurements (gray x's). From Fig. \ref{cardinality}, as the true agents die or birth initially, estimates of the occurrence is accurate. As the number of agents increases, the estimates become less accurate. This is because the GM-PHD filter only uses the first-order statistical moment to propagate the cardinality information of agents \cite{46}. The cardinality distribution is unknown, and it is approximated as a Poisson distribution. For a Poisson distribution, the mean and covariance are equal. Therefore, if there are a larger number of agents in the field, the corresponding covariance of the cardinality distribution is also higher. Although the estimates are less accurate at high cardinality, the individual agents are controlled successfully into a star pattern in the presence of imperfect information. This is shown directly in \ref{snapshot} and \ref{timehis}. As agents die or birth, the control input dies, or births with it, and due to the $L_2^2$ plus quadratic distance, agents are flexible to move into different parts of the formation.
\section{Limitations}
 RFS control for large collaborative swarms provides control solutions that are adaptive to varying swarm size (number of agents) and desired targets in the presence of process, measurement, and cardinality uncertainty, but several limitations currently exist. First, agents are assumed to be identical and unlabeled. Thus, this formulation does not provide control to a specific agent in the field. Secondly, control optimality of the solutions shown in this work are demonstrated through empirical results, but this work lacks theoretical proofs of robustness and optimally. Thirdly, RFS control was applied using a complete topology (centralized control), therefore, computational inefficiencies do exist.  But, the authors believe that RFS-based method can be generalized and extended to decentralized applications. Lastly, no strict collision avoidance methods are applied for this problem. Although repelling behavior between agents exist in the objective function, no strict collision avoidance constraints are applied. Thus, this provides many areas for future work.

\section{Conclusions}
The objective of this paper is to formulate the multi-agent estimation and control background for swarming formations using the Gaussian Mixture Probability Hypothesis Density (GM-PHD) filter and either Quasi-Newton Model Predictive Control (MPC) or Iterative Linear Quadratic Regulator (ILQR) from Random Finite Set (RFS) theory. \textcolor{black}{The RFS formulation is used to control the “concentration” of agents to match a desired
distribution.} By setting up the problem using information divergence to define the distance between the swarm RFS and the desired target configuration, an optimal control problem is found that tracks a linear system with a nonquadratic objection function through the use of Quasi-Newton MPC and ILQR. The results show that the approach can be adaptive to varying swarm size and desired targets.
Lastly, control using RFSs is also applied to the spacecraft relative motion problem by ``closing-the-loop'' with the GM-PHD filter. With the inclusion of process and measurement noise and uncertainty in the large number of agents in the field, a converging control solution is found obtained from estimates from the GM-PHD filter. 
These examples show the benefit of control using RFS by overcoming the curse of dimensionality.

\setcounter{equation}{0}
\renewcommand\theequation{A.\arabic{equation}}
\section*{APPENDIX A: Differential Dynamic Programming}\label{appa}
Finite-horizon LQR control is first discussed to provide the necessary background for DDP discussed afterward.
\subsection*{LQR Finite-Horizon Optimal Control Problem}\label{section11}
The linear quadratic regulator problem is defined by a discrete time-varying system given by
\begin{equation}
{\bf x}_{k+1}=A_k{\bf x}_{k}+B_k{\bf u}_{k}+\boldsymbol{\epsilon}_k,
\end{equation}
where $\boldsymbol{\epsilon}_k$ is Brownian process noise. For the finite horizon $N$, the total cost is calculated from an initial state ${\bf x}_0$ and using the control sequence $U=[{\bf u}_{k},{\bf u}_{k+1},\cdots,{\bf u}_{N-1}]$ applied to the dynamics given by
\begin{equation}\label{costlqr}
J({\bf x}_0,U)=\sum_{k=0}^{N-1}l({\bf x}_k,{\bf u}_k)+l_f({\bf x}_N),
\end{equation}
where $l({\bf x}_k,{\bf u}_k)$ is the running cost and $l_f({\bf x}_N)$ is the terminal cost. The LQR costs are quadratic given by
\begin{equation}
l({\bf x}_k,{\bf u}_k)=\frac{1}{2}\left[ \begin{array}{c} 1 \\ { \bf x}_k \\ \mathbf{u}_k \end{array} \right]^{T} \begin{bmatrix} 0& \mathbf{q}_{k}^{T} & \mathbf{r}_{k}^{T} \\ \mathbf{q}_{k}&Q_{k} & P_{k} \\\mathbf{r}_{k}&P_{k} & R_{k}  \end{bmatrix}\left[ \begin{array}{c} 1\\ { \bf x}_k \\ \mathbf{u}_k \end{array} \right],\hspace{12pt}l_f({\bf x}_N)=\frac{1}{2}{ \bf x}_N ^{T}Q_{N}{ \bf x}_N+{ \bf x}_N ^{T}\mathbf{q}_{N},
\end{equation}
where $\mathbf{q}_{k}$, $\mathbf{r}_{k}$, $Q_{k}$, $R_{k}$, and $P_{k}$ are the running weights (coefficients), and $Q_{N}$ and $\mathbf{q}_{N}$ are the terminal weights. The weight matrices, $Q_{k}$ and $R_{k}$, are positive definite and the block matrix $\begin{bmatrix} Q_{k} & P_{k} \\P_{k} & R_{k}  \end{bmatrix}$ is positive-semidefinite \cite{13}. The costs are substituted into Eq. \eqref{costlqr}, and due to the symmetry in the weight matrices, the total cost is simplified to
\begin{equation}
J({\bf x}_0,U)=\sum_{k=0}^{N-1}{ \bf x}_k ^{T}\mathbf{q}_{k}+{ \bf u}_k^{T}\mathbf{r}_{k}+\frac{1}{2}{ \bf x}_k ^{T}Q_{k}{ \bf x}_k+\frac{1}{2}{ \bf u}_k ^{T}R_{k}{ \bf u}_k +{ \bf u}_k ^{T}P_{k}{ \bf x}_k+\frac{1}{2}{ \bf x}_N ^{T}Q_{N}{ \bf x}_N+{ \bf x}_N ^{T}\mathbf{q}_{N}.
\end{equation}
The optimal control solution is based on minimizing the cost function in terms of the control sequence which is given by
\begin{equation}\label{minJ}
U^*({\bf x}_0)=\argmin_U J({\bf x}_0,U).
\end{equation}
To solve for the optimal control solution given by Eq. \eqref{minJ}, a value iteration method is used. Value iteration is a method that determines the optimal cost-to-go (value) starting at the final time-step and moving backwards in time minimizing the control sequence. Similar to Eq. \eqref{costlqr} and \eqref{minJ}, the cost-to-go and optimal cost-to-go are defined as
\begin{subequations}
\begin{equation}\label{costgo}
J({\bf x}_{k},U_{k})=\sum_{k}^{N-1}l({\bf x}_k,{\bf u}_k)+l_f({\bf x}_N),
\end{equation}
\begin{equation}\label{valuefunction}
V({\bf x}_{k})=\min_{U_{k}} J({\bf x}_{k},U_{k}),
\end{equation}
\end{subequations}
where $U_{k}=[{\bf u}_{k},{\bf u}_{k+1},\cdots,{\bf u}_{N-1}]$. Instead, the cost starts from time-step $k$ instead of $k=0$. 
At a time-step $k$, the optimal cost-to-go function is a quadratic function given by
\begin{equation}\label{quadfun}
V({\bf x}_{k})=\frac{1}{2}{\bf x}_k^{T}S_k{\bf x}_k+{\bf x}_k^{T}\mathbf{s}_k+ c_k,
\end{equation}
where $S_k$, $\mathbf{s}_k$, and $c_k$ are computed backwards in time using the value iteration method. First, the final conditions $S_N=Q_N$, $\mathbf{s}_N=\mathbf{q}_N$, and $c_N=c$ are set. This reduces the minimization of the entire control sequence to just a minimization over a control input at a time-step which is the principle of optimality \cite{31}.  To find the optimal cost-to-go, the Riccati equations are used to propagate the final conditions backwards in time given by
\begin{subequations}
\begin{equation}\label{rit1}
S_k=A_k^{T}S_{k+1}A_k+Q_k-\left(B_k^{T}S_{k+1}A_k+P_k^{T} \right)^{T}\left(B_k^{T}S_{k+1}B_k+R_k \right)^{-1}\left(B_k^{T}S_{k+1}A_k+P_k^{T}\right),
\end{equation}
\begin{equation}\label{rit2}
\mathbf{s}_k=\mathbf{q}_k+A_k^{T}\mathbf{s}_{k+1}+A_k^{T}S_{k+1}\mathbf{g}_k
-\left(B_k^{T}S_{k+1}A_k+P_k^{T}\right)^{T}\left(B_k^{\mathsf{T}}S_{k+1}B_k+R_k\right)^{-1}\left(B_k^{T}S_{k+1}\mathbf{g}_k+B_k^{T}\mathbf{s}_{k+1}+\mathbf{r}_k\right),
\end{equation}
\begin{equation}\label{rit3}
c_k=\mathbf{g}_k^{T}S_{k+1}\mathbf{g}_k+2\mathbf{s}_{k+1}^{T}\mathbf{g}_k+c_{k+1}
-\left(B_k^{T}S_{k+1}\mathbf{g}_k+B_k^{T}\mathbf{s}_{k+1}+\mathbf{r}_k\right)^{T}\left(B_k^{\mathsf{T}}S_{k+1}B_k+R_k\right)^{-1}\left(B_k^{T}S_{k+1}\mathbf{g}_k+B_k^{T}\mathbf{s}_{k+1}+\mathbf{r}_k\right).
\end{equation}
\end{subequations}
Using the Ricatti solution, the optimal control policy is in the affine form
\begin{equation}
\mathbf{u}_k({\bf x}_k)=K_k{\bf x}_k+\mathbf{l}_k,
\end{equation}
where the controller, $K_k$, and controller offset is given by
\begin{subequations}
\begin{equation}\label{controlk}
K_k=-(R_k+B_k^{T}S_{k+1}B_k)^{-1}(B_k^{T}S_{k+1}A_k+P_k^{T}),
\end{equation}
\begin{equation}\label{controloffset}
\mathbf{l}_k=-(R_k+B_k^{T}S_{k+1}B_k)^{-1}(B_k^{T}S_{k+1}\mathbf{g}_k+B_k^{T}\mathbf{s}_{k+1}+\mathbf{r}_k).
\end{equation}
\end{subequations}

This optimal solution to the LQR problem works for linear dynamics and quadratic cost functions, but unfortunately, the objective function specified for the swarm problem is nonquadratic. Fortunately, differential dynamic programming can be used for nonlinear dynamics and nonquadratic local cost functions.

\subsection*{The Differential Dynamic Programming Problem}\label{section22}
The DDP approach to solving nonlinear and nonquadratic equations uses a similar process as the LQR solution, but a second-order approximation of the dynamics and objective function are obtained for value iteration and the solution is iterated to increasingly get better approximations of the optimal trajectory of the system. Note that if linear dynamics are used, the ILQR formulation is obtained \citep{32,11}. Since the results are produced by a linear system, both the DDP and ILQR terms can be used interchangeably. The following discussion on DDP follows closely to that of Tassa \cite{32,tassa2012synthesis}. The general nonlinear discrete-time dynamics is given by
\begin{equation}\label{nonlineq}
{\bf x}_{k+1}=f({\bf x}_k,{\bf u}_k),
\end{equation}
where the state at the next time-step, ${\bf x}_{k+1}$, is a function of the current state, ${\bf x}_k$, and control input ${\bf u}_k$. The cost function is in the form of Eq. \eqref{costlqr}, but the costs are nonquadratic. The solution to the optimal control problem is Eq. \eqref{minJ}. Similarly, the cost-to-go and the optimal cost-to-go function are defined by Eq. \eqref{costgo} and Eq. \eqref{valuefunction} respectively. Given the terminal condition $V({\bf x}_{N})=l_f({\bf x}_N)$, the optimal cost-to-go obtained from the  principle of optimality is
\begin{equation}\label{valuenon}
V({\bf x}_{k})=\min_{\mathbf{u}_{k}} \left( l({\bf x}_k,{\bf u}_k)+V({\bf x}_{k+1})\right),
\end{equation}
which minimizes over the control at a time-step and solved through time by a backwards pass (value iteration).

\subsubsection{Backward Pass}
The first step in the backward pass (value iteration) is to determine a value function that is quadratic. The argument in Eq. \eqref{valuenon} is taken as a function of small perturbations around the state ($\delta\mathbf{x}_k$) and control input ($\delta\mathbf{u}_k$), and it is quadratized through a second order Taylor series expansion given by

\begin{equation}\label{valueexpand}
\begin{aligned}
Q(\delta\mathbf{x},\delta\mathbf{u})&=l(\mathbf{x}_k+\delta\mathbf{x}_k,\mathbf{u}_k+\delta\mathbf{u})_k-l(\mathbf{x},\mathbf{u})+V({\bf x}_{k+1}+\delta\mathbf{x}_{k+1})-V({\bf x}_{k+1}),\\
&\approx\frac{1}{2}\left[ \begin{array}{c} 1 \\  \delta\mathbf{x}_k \\ \delta\mathbf{u}_k \end{array} \right]^{T} \begin{bmatrix} 0& Q_{x}^{T} & Q_{u}^{T} \\ Q_{x}&Q_{xx} & Q_{xu} \\Q_{u}&Q_{ux} & Q_{uu}  \end{bmatrix}\left[ \begin{array}{c} 1\\  \delta\mathbf{x}_k \\ \delta\mathbf{u}_k \end{array} \right],
\end{aligned}
\end{equation}
where  $Q_x$, $Q_u$, $Q_{xx}$, $Q_{xu}$, and $Q_{uu}$ are the running coefficients (weights) of the quadratized value function at a certain time-step. Note, in the standard formulation, the time-step $k$ is dropped for these equations. Any primes denote the next time-step. The equations for the running weights are given by
\begin{subequations}\label{quadapproximations}
\begin{equation}
Q_x= l_x+ f_x^{T}V_x',
\end{equation}
\begin{equation}
Q_u= l_u+ f_u^{T}V_x',
\end{equation}
\begin{equation}
Q_{xx}= l_{xx}+ f_x^{T}V_{xx}'{ f}_x+V_x' f_{xx},
\end{equation}
\begin{equation}
Q_{uu}= l_{uu}+ f_u^{T}V_{xx}'{ f}_u+V_x' f_{uu},
\end{equation}
\begin{equation}
Q_{ux}= l_{ux}+ f_u^{T}V_{xx}'{ f}_x+V_x' f_{ux},
\end{equation}
\end{subequations}
where $l_x$, $l_u$, $l_{xx}$, $l_{uu}$, and $l_{ux}$ are the gradients and Hessians of the cost function, $f_x$, $f_u$, $f_{xx}$, $f_{uu}$ $f_{ux}$ are the gradients and Hessians of the nonlinear dynamics, and $V_x'$, and $V_{xx}'$ are the gradient and Hessian of the value function. For the ILQR formulation, the gradients and Hessians for an LTV system (which is the model used for RFS control in Section \ref{dynamicmodelsec}) are trivial, but for the DDP formulation, the gradients and Hessians for the nonlinear dynamics must be computed. By using this quadratic approximation, the minimum in terms of $\delta\mathbf{u}$ is found using
\begin{equation}\label{valueperturb}
\delta\mathbf{u}=\argmin_{\delta\mathbf{u}}Q\left(\delta\mathbf{x},\delta\mathbf{u}\right)=-Q_{uu}^{-1}\left( Q_u+Q_{ux}\delta\mathbf{x}\right),
\end{equation}
which provides local feedback and feed-forward gains of 
\begin{subequations}\label{controllerKk}
\begin{equation}
K=-Q_{uu}^{-1}Q_{ux},
\end{equation}
\begin{equation}
\mathbf{k}=-Q_{uu}^{-1}Q_{u},
\end{equation}
\end{subequations}
respectively. The locally optimal controller is substituted back into Eq. \eqref{valueexpand} to get the optimal value given by
\begin{subequations}\label{vxvxx}
\begin{equation}
\Delta V=-\frac{1}{2}\mathbf{k}^{T}Q_{uu}\mathbf{k},
\end{equation}
\begin{equation}
V_x=Q_x-K^{T}Q_{uu}\mathbf{k},
\end{equation}
\begin{equation}
V_{xx}=Q_{xx}-K^{T}Q_{uu}K,
\end{equation}
\end{subequations}
so the value can be propagated backwards in time to find new locally optimal solutions to the value function.
\subsubsection{Forward Pass}
By continually computing the quadratic approximations in Eq. \eqref{quadapproximations}, local controller in Eq. \eqref{controllerKk}, and the new values in Eq. \eqref{vxvxx} backwards in time from the terminal condition $V({\bf x}_{N})=l_f({\bf x}_N)$, the updated trajectory can be found through a forward pass given by
\begin{subequations}\label{forwardpass}
\begin{equation}
\mathbf{\hat{x}}_0=\mathbf{x}_0,
\end{equation}
\begin{equation}
\mathbf{\hat{u}}_k=\mathbf{u}_k+\mathbf{k}_k+K_k\left(\mathbf{\hat{x}}_k-\mathbf{x}_k\right),
\end{equation}
\begin{equation}\label{hatnewdynamics}
{\bf \hat{x}}_{k+1}=f({\bf \hat{x}}_k,{\bf \hat{u}}_k).
\end{equation}
\end{subequations}
where $\mathbf{\hat{x}}_k$ and $\mathbf{\hat{u}}_k$ consists of the state and control input at a time-step of the new trajectory $\left(\mathbf{\hat{X}}, \mathbf{\hat{U}} \right)$.
This composes one iteration of DDP. If the cost of the new trajectory, $\left(\mathbf{\hat{X}}, \mathbf{\hat{U}} \right)$, is less than the cost of the old trajectory, $\left(\mathbf{X}, \mathbf{U} \right)$, then $\mathbf{X}= \mathbf{\hat{X}}$ and $\mathbf{U}= \mathbf{\hat{U}} $ are set, and the algorithm is ran again until a convergence threshold is met between the old and new costs. 

\subsubsection{Regularization via the Levenberg-Marquardt Heuristic}
If the cost of the new trajectory is greater than the cost of the old trajectory, the iteration has not provided a better solution. To circumvent this issue, the Hessian is regularized. This is called the Levenberg-Marquardt heuristic. The control sequence that is calculated in DDP is computed like a Newton optimization which uses second order information (curvature information) on top of the first order information (gradient information) \cite{33}. By including second order curvature to the update, optimization can occur faster, but this relies on the fact that the Hessian is positive definite and an accurate quadratic model. If the control update is not improving (for a non-positive definite Hessian and inaccurate quadratic model), the Levenberg-Marquardt heuristic uses less curvature information and more on the gradient information. This regularization is added to the Hessian of the control cost given by
\begin{equation}
\tilde{Q}_{uu}=Q_{uu}+\mu I_m,
\end{equation}
where $\tilde{Q}_{uu}$ is the regularized control cost Hessian, $\mu$ is the Levenberg-Marquardt parameter, and $I_m$ is the identity matrix that is the size of the control input vector \cite{34}. This allows for the increase or decrease of curvature information in the optimization by adding a quadratic cost around the current control input. Unfortunately, adding this regularization term can have different effects at different time-steps using the same control perturbation based on a changing $f_u$ in the linearized dynamics. By increasing $\mu\rightarrow\infty$, the $\mathbf{k}$ and $K$ gains become very small due to the $\tilde{Q}_{uu}^{-1}$ term. Therefore, the regularization term is improved by penalizing the states instead of the control inputs which are given by
\begin{subequations}\label{regqk}
\begin{equation}\label{regquu}
\tilde{Q}_{uu}=l_{uu}+ f_u^{T}\left(V_{xx}'+\mu I_n\right){ f}_u+V_x' f_{uu},
\end{equation}
\begin{equation}
\tilde{Q}_{ux}= l_{ux}+ f_u^{T}\left(V_{xx}'+\mu I_n\right){ f}_x+V_x' f_{ux},
\end{equation}
\begin{equation}
K=-\tilde{Q}_{uu}^{-1}\tilde{Q}_{ux},
\end{equation}
\begin{equation}
\mathbf{k}=-\tilde{Q}_{uu}^{-1}Q_{u},
\end{equation}
\end{subequations}
where $I_n$ is the identity matrix that is the size of the state vector. The $\mu$ parameter is placed on the state instead of the control input. For this method, the regularization term is directly incorporated with $f_u$, and the feedback gains, $\mathbf{k}$ and $K$, do not disappear as $\mu\rightarrow\infty$. Instead, the new $\mathbf{k}$ and $K$ values bring the new trajectory closer to the old one. For the implementation of the $\mu$ term, three requirements should be followed. If reaching the minimum is accurate, the $\mu$ should become zero in order to obtain faster convergence due to the second order optimization term. If a non-positive definite $\tilde{Q}_{uu}$ is found, the backward pass should be restarted with a larger $\mu$. The last requirement is that when a $\mu>0$ is needed, the smallest $\mu$ should be used that allows the $\tilde{Q}_{uu}$ to be positive definite. Therefore, more of the second order information can be used to provide faster convergence than gradient descent. The specific algorithm is found in \cite{tassa2012synthesis}.

Eq. \eqref{vxvxx} must also be modified based on regularization added in Eq. \eqref{regquu} \cite{11}. Eq. \eqref{vxvxx} was originally derived using Eqs. \eqref{valueexpand} and \eqref{valueperturb}, but using the new regularized terms in Eq. \eqref{regquu} creates error. Therefore, the modified values at a time-step $k$ are
\begin{subequations}\label{regvxvxx}
\begin{equation}
\Delta V=\frac{1}{2}\mathbf{k}^{T}Q_{uu}\mathbf{k}+\mathbf{k}^{T}Q_u,
\end{equation}
\begin{equation}
V_x=Q_x+K^{T}Q_{uu}\mathbf{k}+K^{T}Q_u+Q_{ux}^{T}\mathbf{k},
\end{equation}
\begin{equation}
V_{xx}=Q_{xx}+K^{T}Q_{uu}K+K^{T}Q_{ux}+Q_{ux}^{T}K.
\end{equation}
\end{subequations}
The regularization terms create a faster and more accurate solution to the backwards pass of the DDP solution.
\subsubsection{Forward Pass Line Search}
Regularization of the forward pass can improve convergence and performance of the DDP algorithm. For linear time-varying systems, one iteration provides a minimal solution after one iteration. This is not the case for general nonlinear systems. Since nonlinear systems are approximated by a Taylor series expansion, there may be regions in the new DDP trajectory that are not valid about the nonlinear model. This may lead to divergence and have a larger cost function than the old trajectory. To fix this issue, a backtracking line-search parameter is introduced in the control update equation given by
\begin{equation}
\mathbf{\hat{u}}_k=\mathbf{u}_k+\alpha\mathbf{k}_k+K_k\left(\mathbf{\hat{x}}_k-\mathbf{x}_k\right),
\end{equation}
where $\alpha$ is set to $\alpha=1$ at the start of the forward pass. Then the expected cost reduction is considered using
\begin{equation}\label{alphacontrol}
\Delta J(\alpha)=\alpha\sum_{k=0}^{N-1}\mathbf{k}(k)^{T}Q_u(k)+\frac{\alpha^2}{2}\sum_{k=0}^{N-1}\mathbf{k}^{T}(k)Q_{uu}(k)\mathbf{k}(k).
\end{equation}
A ratio $z$ is determined using the actual and expected cost reduction given by
\begin{equation}
z=\frac{\left( J({\bf x}_0,U)-J(\hat{{\bf x}}_0,\hat{U})\right)}{\Delta J(\alpha)},
\end{equation}
where $J({\bf x}_0,U)$ and $J(\hat{{\bf x}}_0,\hat{U})$ are the old and new cost respectively. The control update is accepted if the condition,
\begin{equation}\label{reductioncondition}
0<c_1<z,
\end{equation}
is met where $c_1$ is a parameter set by the user. The $c_1$ is usually set close to zero. If the condition is not met, the forward pass is restarted with a smaller $\alpha$ value which means that the new trajectory strayed farther than the system's region of validity. By using the $\alpha$ line search parameter, convergence can be achieved for nonlinear systems by iteratively deceasing $\alpha$ to obtain a cost reduction.
\subsubsection{DDP Summary}
A DDP iteration can be summarized in four steps. First, an initial rollout of the nonlinear dynamics given by Eq. \eqref{nonlineq} is integrated over time for a given control sequence $U$. If there is no good initialization of the control sequence, the control sequence can be set to $U=0$. After the initial rollout, the derivatives of the cost function and nonlinear dynamics used in Eq. \eqref{quadapproximations} are found. The derivatives are used in the third step which is to determine local control solutions using a backward pass. Using the terminal condition, $V({\bf x}_{N})=l_f({\bf x}_N)$, local control solutions are found by iterating Eq. \eqref{quadapproximations}, \eqref{regqk}, and \eqref{regvxvxx} backwards at each time-step. When a non-positive definite $\tilde{Q}_{uu}$ is found, increase the regularization parameter $\mu$ and restart the backward pass. Once a local optimal policy is found, $\alpha$ is set to $\alpha=1$, and Eqs. \eqref{hatnewdynamics} and \eqref{alphacontrol} are propagated forward in time. If the integration diverged or cost reduction condition in Eq. \eqref{reductioncondition} was not met, the forward pass is restarted with a smaller $\alpha$.

\setcounter{equation}{0}
\renewcommand\theequation{B.\arabic{equation}}
\section*{APPENDIX B: Receding Horizon Control using the RFS Formulation}\label{appb}

An optimal solution, ${\bf u}$, can also be obtained in a real-time computational sense by minimizing the objective, Eq. \eqref{modifiedL2}, by reducing the finite horizon to a computational manageable prediction horzion using MPC or receding horizon control \cite{15}. Conceptually, at a time k, the knowledge of the system model is used to derive a sequence ${\bf u}(k|k), {\bf u}(k+1|k), {\bf u}(k+2|k),\cdots, {\bf u}(k+T_p|k)$ where $T_p$ is the finite prediction horizon from the current state ${\bf x}(k)$ \cite{14}. With the input sequence, the state is moved forward in time by the control horizon, $T_c$; usually one time-step. Then the same strategy is repeated for time $k+1$. The finite prediction horizon, $T_p$, can be chosen to be either small or large. As $T_p$ increases, the degrees of freedom in the optimization increase which can slow down the algorithm considerably, even though more of the future reference trajectory would be useful to bring the output closer to the reference. With a smaller $T_p$, the computation time will be faster, but the optimization may be more suboptimal. Thus, the swarm may not converge to the desired configuration.

For the RFS control formulation, a ${\bf u}$ that controls the swarm intensities through their statistics (mean and covariance) is found by minimizing the objective as given by Eq. \eqref{overalleqn} and \eqref{qwer}. This can be done by using MPC via the Quasi-Newton method or DDP. DDP is able to determine an optimal solution for nonlinear equations of motion and a nonquadratic cost function through an iterative process of finding the solution involving second-order approximations of the dynamics and the objective function. The dynamical systems used in the results are linear, thus, DDP can be formed as its variant, ILQR. For the Quasi-Newton method, the optimal control input ${\bf u}$ is found using MATLAB's fminunc solver \cite{30}. Note that MPC via DDP or the Quasi-Newton method are both closed-loop control methods in terms of the statistics (mean and covariance) of the system. 
\section*{ACKNOWLEDGMENT}
The authors wish to acknowledge support by the National Aeronautics and Space Administration under Contract Number NNX16CP45P issued through the NASA STTR Program by the Jet Propulsion Laboratory (JPL) and led by Amir Rahmani at JPL. This work was also supported in part by ONR Code 321 and in part by NSF ATD Grant 1738010. The authors wish to acknowledge useful conversations related to satellite technologies with Chuck Hisamoto, Vaughn Weirens, and Suneel Sheikh of ASTER Labs, Inc. The authors wish to acknowledge Andrew Akerson, a graduate student at the California Institute of Technology, who enabled useful notation for the mathematical theory used in the paper. Lastly, the authors wish to acknowledge Piyush M. Mehta, Assistant Professor at West Virginia University, for providing useful conversations in structuring and optimizing code.
\bibliography{sample}

\end{document}